\documentclass[journal,10pt,letterpaper]{IEEEtran}

\usepackage{epstopdf}
\usepackage{graphicx}
\usepackage{stmaryrd}
\usepackage{amsmath,amsthm,amssymb}
\usepackage{multirow,url}
\usepackage{color,cite}
\usepackage{algorithm}
\usepackage{algorithmicx}
\usepackage{algpseudocode}
\usepackage{setspace}
\usepackage{rotating}
\usepackage{bm}

\newtheorem{cor}{Corollary}

\newtheorem{thm}{Theorem}
\newtheorem{defn}{Definition}

\newtheorem{pp}{Proposition}

\ifodd 0
\newcommand{\com}[1]{\textbf{\color{red} (COMMENT: #1)}} 
\newcommand{\comg}[1]{\textbf{\color{green} (COMMENT: #1)}}
\newcommand{\response}[1]{\textbf{\color{magenta} (RESPONSE: #1)}} 
\else

\newcommand{\com}[1]{}
\newcommand{\comg}[1]{}
\newcommand{\response}[1]{}
\fi

\begin{document}


\title{Exploiting Social Tie Structure for Cooperative Wireless Networking: A Social Group Utility Maximization Framework}

\author{Xu Chen, \emph{Member, IEEE},  Xiaowen Gong, \emph{Student Member, IEEE}, Lei Yang, \emph{Member, IEEE}, and Junshan Zhang, \emph{Fellow, IEEE}
}

\maketitle

\pagestyle{empty}
\thispagestyle{empty}
\allowdisplaybreaks

\begin{abstract}
In this paper, we develop a social group utility maximization (SGUM) framework for cooperative wireless networking that takes into account both social
relationships and physical coupling among users. Specifically, instead of maximizing its individual utility or the overall network utility, each user
aims to maximize its social group utility that hinges heavily on its social tie structure with other users. We show that
this framework provides rich modeling flexibility and spans the continuum between non-cooperative game and network utility maximization (NUM) -- two traditionally disjoint paradigms for network optimization. Based on this framework, we study three important applications of SGUM, in database assisted spectrum access, power control, and random access control, respectively. For the case of database assisted spectrum access, we show that the SGUM game is a potential game and always admits a socially-aware Nash equilibrium (SNE). To overcome the suboptimality associated with the asynchronous best response update approach, we develop a randomized distributed spectrum access algorithm that can asymptotically converge to the optimal SNE with high probability. Using spectral gap analysis and path coupling argument, we derive upper bounds on the convergence time in the Glauber dynamics and also quantify the trade-off between the performance and convergence time of the algorithm. We further show that the performance gap of SNE by the algorithm from the NUM solution decreases as the strength of social ties among users increases and the performance gap is zero when the strengths of social ties among users reach the maximum values. For the cases of power control and random access control, we show that there exists a unique SNE. Furthermore, as the strength of social ties increases from the minimum to the maximum, a player's SNE strategy (i.e., access probability or transmit power) migrates from the Nash equilibrium strategy in a standard non-cooperative game to the socially-optimal strategy in network utility maximization. Numerical results corroborate that the SGUM solutions can achieve superior performance using real social data trace. Furthermore, we show that the SGUM framework can be generalized to take into account both positive and negative social ties among users. The generalized SGUM framework also encompasses the zero-sum game as a special case and can be a useful tool for studying network security problems.
\end{abstract}


\section{Introduction}
\begin{figure}[tt]
\begin{center}
\includegraphics[scale=0.5]{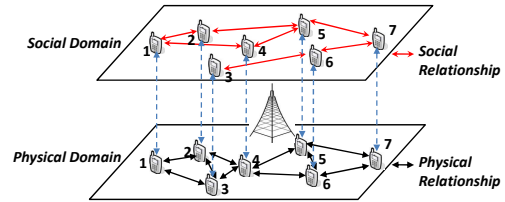}
\caption{\label{fig:PhySocial}An illustration of the social group utility maximization framework. In the physical domain, users have different physical coupling subject to physical relationships (e.g., interference). In the social domain, users have heterogeneous social coupling due to the social ties among users.}
\end{center}
\end{figure}

Mobile networks have been projected to continue growing rapidly in the foreseeable future. For example, mobile phone shipments are projected to reach $11.5$ billions in $2019$ \cite{Cisco} and mobile data traffic is predicted to grow further by over $100$ times in the next ten years \cite{Cisco}. Different from other networks (e.g., sensor networks), a distinctive characteristic of mobile networks is that mobile devices are carried and
operated by human beings. With the explosive growth of online social networks such as Facebook and Twitter, more and more people are actively involved in online social interactions, and social relationships among people are hence extensively broadened and significantly enhanced. Then it is natural to ask whether ``social relationships" among
mobile users can impact the communications and interactions among their devices, and if yes, how the social structure can
be cleverly leveraged?  With this insight, as illustrated in Figure \ref{fig:PhySocial}, we view a mobile network as
an overlay/underlay system where a ``virtual social network'' overlays the physical communication network (the  ``social network'' is virtual, in the sense that the social tie structure therein results from  existing human relationship and  online social networks), and leverage the intrinsic social tie structure among mobile users  to facilitate cooperative wireless networking.

\begin{figure}[tt]
\begin{center}
\includegraphics[scale=0.6]{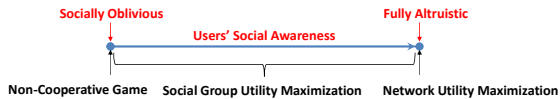}
\caption{\label{fig:Spectrum}Network optimization with different users' social awareness levels}
\end{center}
\end{figure}

With this motivation, we advocate a novel \emph{social group utility maximization} (SGUM) framework that takes into account both the
users' social relationships and physical coupling. As illustrated in Figure \ref{fig:PhySocial}, a key observation
is that users are coupled not only in the physical domain due to the physical relationship (e.g., interference), and but also in the social domain due to the social ties among them. It would be a win-win case for users to help those users having social ties with them. Specifically, we cast the distributed decision making problem for cooperative networking among users as a SGUM game, where each user maximizes its social group utility, defined as the sum of its own individual utility and the weighted sum of the utilities of other users having social tie with it. Interestingly, as illustrated in Figure \ref{fig:Spectrum}, the SGUM framework can bridge the gap between non-cooperative game and network utility maximization -- two traditionally disjoint paradigms for network optimization.  These two paradigms are captured under the proposed framework as two special cases where no social tie exists among users (i.e., users are socially oblivious) and all users are connected by the strongest social ties (i.e., users are fully altruistic), respectively.

To get a more concrete sense of the SGUM framework, in this paper we study three important network applications, namely, database assisted spectrum access, power control, and random access control. For the case of database assisted spectrum access, we prove that the SGUM game is a potential game and always admits a socially-aware Nash equilibrium (SNE). Moreover, we show that the potential function of the game exhibits a nice structure that can be decomposed into two parts, capturing the impact of the physical coupling and social coupling in spectrum access, respectively. To overcome the drawback that
the asynchronous best response update approach would achieve a suboptimal SNE, we design a randomized distributed spectrum access algorithm that can achieve the optimal SNE (that maximizes the potential function) of the SGUM game for database assisted spectrum access. Using Markov chain uniformization and spectral analysis, we derive an upper bound on the convergence time in the Glauber dynamics and also quantify the trade-off between the performance and convergence time of the algorithm. We further derive the upper-bound of the performance gap of the SNE from the NUM solution, and show that the upper-bound of the performance gap decreases as the strength of social
ties among users increases and the performance gap is zero when the strengths of social ties among users reach the maximum values.

For the cases of power control and random access control, we show that there exists a unique SNE. Furthermore, as the strength of social ties increases from the minimum (selfish) to the maximum (fully altruistic), a player's SNE strategy (i.e., access probability or transmit power) migrates from the Nash equilibrium strategy in a standard non-cooperative game to the social-optimal strategy in network utility maximization.

We should emphasize that the SGUM framework highlights the interplay between the physical coupling subject to users' physical relationships and the social coupling due to the social ties among users. The SGUM framework spans the continuum between non-cooperative game and network utility maximization -- two extreme paradigms based on drastically different assumptions that users are selfish and altruistic, respectively, and hence can provide rich flexibility for modeling cooperative networking problems.

The rest of the paper is organized as follows. We discuss the related work in Section \ref{ReWork} and introduce the SGUM framework in Section \ref{sec:framework}. We apply the framework to the applications of database assisted spectrum access, power control, random access control in Sections \ref{sc:power}, \ref{sc:random}, and \ref{sec:System-Model}, respectively. We discuss the possible generalization of the SGUM framework in Section \ref{extension}, and finally conclude the paper in Section \ref{conclusion}.

\section{Related Work}\label{ReWork}
Game theory has been extensively used for a variety of networking applications \cite{Altman06}. A standard non-cooperative game assumes that all users are \emph{selfish}, i.e., each user  cares about only its own welfare and behaves without regard to its impact on other users. Along a different line, network utility maximization has been extensively studied for resource allocation problems of various networks \cite{Palomar06}, where all users are assumed to have the same objective of maximizing the total welfare of all users (i.e., network utility). However, all these paradigms cannot capture a mobile social network where users have diverse social relationships, which is a primary objective of the SGUM framework under consideration. In fact, the SGUM framework provides rich modeling flexibility by spanning the continuum between these traditionally disjoint paradigms.

Although there exists a significant body of work on non-cooperative game and network utility maximization, very little attention has been paid to the continuum between these extreme paradigms, especially in the context of mobile social networks. Recent works~\cite{chen2008altruism,hoefer2009altruism} have studied the impact of altruistic behavior in a routing game. \cite{Kesidis10} has recently investigated a random access game between two symmetrically altruistic users. \cite{ashlagi2008social} introduces different approaches to integrate user's social awareness into the payoff function design, including the cases that a user cares about only its best/worst performing friend or the friends within some specific communities (e.g., the cliques by partitioning the social graph). Moreover, \cite{ashlagi2008social}  considers only the  binary social setting (i.e., friend or not). In contrast, in the SGUM framework we consider a general weighted social graph and define the social group utility function as the weighted sum of the individual utilities of the social neighbors with the weights proportional to the social tie strength among the users. \cite{hoefer2012social} considers a similar socially-aware payoff function as that in the SGUM framework. Nevertheless, \cite{hoefer2012social} focuses on the symmetric social network and mainly considers the congestion game in which the payoff of a player depends on the number of players choosing the same strategy. In contrast, the payoff of a user in the SGUM games in mobile networks depends on a variety of factors including interference relationships and the specific interacting user set. We should emphasize that a key objective of this paper to develop a general modeling framework for exploiting the social relationships among  users for cooperative wireless networking design. To this end, we propose the SGUM framework and introduce both physical and social graphs to model the physical and social couplings among users in both physical and social domains, respectively. We also highlight its connections with non-cooperative game and NUM, two widely adopted modeling frameworks in wireless networks. These features are missing in \cite{ashlagi2008social,hoefer2012social}.


The social aspect is now becoming a new and important dimension for
communication system design \cite{kayastha2011applications,chen2014exploiting}. As the development
of online and mobile social networks such as Facebook and Twitter, more and more
real-world data and traces of human social interactions are being generated. This enables researchers
and engineers to observe, analyze, and incorporate the social
factors into engineering system design in a way never previously
possible \cite{kayastha2011applications}.  A channel recommendation system based on cooperative social interactions is developed for dynamic spectrum access in \cite{chen2012adaptive}. Gao \emph{et al.} in \cite{gao2009multicasting} exploited social structures such as social
community to design efficient data forwarding
and routing algorithms in delay tolerant networks. Hui \emph{et al.} in \cite{hui2011bubble} used the social betweenness and centrality as the forwarding metric. Costa \emph{et al.} in \cite{costa2008socially} proposed predictions based on metrics of social interaction to identify the best information carriers for content publish-subscribe.

\section{\label{sec:framework}A Social Group Utility Maximization (SGUM) Framework}
In this section we introduce the SGUM framework for cooperative networking. As illustrated in Figure \ref{fig:PhySocial}, the framework can be projected onto two domains: the physical domain and the social domain. In the physical domain, different wireless users have different physical coupling due to their heterogeneous physical relationships (e.g., interference). In the social domain, different users have different social coupling due to their intrinsic social ties. We next discuss both physical and social
domains in detail.

\subsection{A Physical Network Graph Model}

We consider a set of wireless users $\mathcal{N}=\{1,2,...,N\}$ where $N$ is the total number of users. We denote the set of feasible strategies for each user $n\in\mathcal{N}$ as $\mathcal{X}_{n}$. For instance, a strategy $x\in\mathcal{X}_{n}$ can be choosing either a channel or a power level for wireless transmissions. Subject to heterogeneous physical constraints, the strategy set $\mathcal{X}_{n}$ can be user-specific. For example, the strategy set $\mathcal{X}_{n}$ can be a set of feasible relay users that are in vicinity of user $n$ for cooperative communication.


To capture the physical coupling among the users in the physical domain, we introduce the \emph{physical graph} $\mathcal{G}^{p}=\{\mathcal{N},\mathcal{E}^{p}\}$ (see Figure
\ref{fig:PhySocial} for example). Here the set of users $\mathcal{N}$
is the vertex set, and $\mathcal{E}^{p}\triangleq\{(n,m):e_{nm}^{p}=1,\forall n,m\in\mathcal{N}\}$
is the edge set where $e_{nm}^{p}=1$ if and only if users $n$
and $m$ have physical coupling, i.e.,  users $n$ and $m$ can affect each other's payoff by taking some actions. For example, two users have the physical coupling if they can cause interference to each other when using the same channel for data transmission. We also denote
the set of users that have physical coupling with user $n$ as
$\mathcal{N}_{n}^{p}\triangleq\{m\in\mathcal{N}:e_{nm}^{p}=1\}$.

Let $\boldsymbol{x}=(x_{1},...,x_{N})\in\prod_{n=1}^{N}\mathcal{X}_{n}$ be the strategy profile of all users. Given the strategy
profile $\boldsymbol{x}$, the individual utility function of user $n$ is denoted as $U_{n}(\boldsymbol{x})$, which represents the payoff of user $n$, accounting for the physical coupling among users. For example, $U_{n}(\boldsymbol{x})$ can be the achieved data rate or the satisfaction of quality of service (QoS) requirement of user $n$ under the strategy profile $\boldsymbol{x}$. Note that in general the underlying physical graph plays a critical role in determining the individual utility $U_{n}(\boldsymbol{x})$. For example, users' achieved data rates are determined by the interference graph and channel quality.

\subsection{A Social Network Graph Model}

We next introduce the social graph model to describe the social ties among users. The underlying rationale of considering social ties is that the hand-held devices are carried by human beings and the knowledge of human social
ties can be utilized to enhance the performance of cooperative networking.

Specifically, we introduce the \emph{social graph} $\mathcal{G}^{s}=\{\mathcal{N},\mathcal{E}^{s}\}$
to model the social ties among the users. Here the vertex set
is the same as the user set $\mathcal{N}$ and the edge set is given
as $\mathcal{E}^{s}=\{(n,m):e_{nm}^{s}=1,\forall n,m\in\mathcal{N}\}$
where $e_{nm}^{s}=1$ if and only if users $n$ and $m$ have
social tie between each other, which can be kinship, friendship, or
colleague relationship between two users. Furthermore, for a pair of users $n$ and $m$ who have a social
edge between them on the social graph, we formalize
the strength of social tie as $w_{nm}\in[0,1]$,
with a higher value of $w_{nm}$ being a stronger social tie. We define user $n$'s \emph{social group} $\mathcal{N}_{n}^{s}$ as
the set of users that have social ties with user $n$, i.e.,
$\mathcal{N}_{n}^{s}=\{m:e_{nm}^{s}=1,\forall m\in\mathcal{N}\}$.

To identify the social relationships among users, two users can locally carry out the
identification process through the proximity communications (e.g., using Bluetooth/WiFi-Direct/Device-To-Device communications).
Two users can detect their social relationship by
carrying out the ``matching" process to identify the common
social features among them. For example, two users can
match their mobile phones' contact books. If they have the
phone numbers of each other or many of their phone numbers
are the same, then it is very likely that they know each
other. As another example, two users can match their
home and working addresses and identify whether they are
neighbors or colleagues. Furthermore, two users can
detect the social relationship among them by accessing to
the online social networks such as Facebook and Twitter.
For example, Facebook has exposed access to their social
graph including the objects of friends, events, groups, profile
information, and photos. Any authenticated Facebook user
can have access to these information through the OpenGraph
API \cite{API2010}. Based on the identified social relationships, a user $n$ can then specify the social tie strength $w_{nm}$ towards its social neighbors $m$ in the social group  $\mathcal{N}_{n}^{s}$.

Based on the physical and social graph models above, users are coupled in the physical domain due to the physical relationships, and also coupled in the social domain due to the social ties among them. It would be a win-win case for users to help those users having social ties with them. With this insight, we further define the \emph{social group utility} of each user $n$ as\begin{equation}
S_{n}(\boldsymbol{x})=U_{n}(\boldsymbol{x})+\sum_{m\in\mathcal{N}_{n}^{s}}w_{nm}U_{m}(\boldsymbol{x}).\label{eq:SocialUtility0}
\end{equation}
\emph{It follows that the social group utility of each user consists of two parts: 1) its own individual utility and 2) the
weighted sum of the individual utilities of other users having social ties with it.} In a nutshell,
the social group utility function captures the feature that each user is socially-aware and cares about the users having social ties with it.

\subsection{Social Group Utility Maximization Game}

We next consider the distributed decision making problem among the users, aiming to maximize their social group utilities. Let $x_{-n}=(x_{1},...,x_{n-1},x_{n+1},...,x_{N})$ be the set of strategies chosen by all other users except user $n$. Given the other users' strategies $x_{-n}$, user $n$
wants to choose a strategy $x_{n}\in\mathcal{X}_{n}$ to maximize
its social group utility, i.e.,
\[
\max_{x_{n}\in\mathcal{X}_{n}}S_{n}(x_{n},x_{-n}),\forall n\in\mathcal{N}.
\]
The distributed nature of the problem above naturally
leads to a formulation based on game theory such that each user aims to maximize
its social group utility.  We thus formulate the decision making problem among the users as a strategic game $\Gamma=(\mathcal{N},\{\mathcal{X}_{n}\}_{n\in\mathcal{N}},\{S_{n}\}_{n\in\mathcal{N}})$,
where the set of users $\mathcal{N}$ is the set of players,
$\mathcal{X}_{n}$ is the set of strategies
for each user $n$, and the social group utility function
$S_{n}$ of each user $n$ is the payoff function of player $n$.  In the sequel, we call the game $\Gamma$ as the SGUM game. We next introduce the concept of \emph{socially-aware Nash equilibrium} (SNE).

\begin{defn}
A strategy profile  $\boldsymbol{x}^{*}=(x_{1}^{*},...,x_{N}^{*})$ is a socially-aware Nash equilibrium of the SGUM game if no
player can improve its social group utility by unilaterally changing its strategy, i.e.,
\[
x_{n}^{*}=\arg\max_{x_{n}\in\mathcal{X}_{n}}S_{n}(x_{n},x_{-n}),\forall n\in\mathcal{N}.
\]
\end{defn}


It is worth noting that under different social graphs, the proposed SGUM game formulation can provide rich flexibility for modeling the network optimization problem (as illustrated in Figure \ref{fig:Spectrum}). When the social graph consists of isolated nodes with $w_{nm}=0$ for any $n,m\in\mathcal{N}$ (i.e., all users are selfish), the SGUM game degenerates to the non-cooperative game formulation. When the social graph is fully meshed with edge weight $w_{nm}=1$ for any $n,m\in\mathcal{N}$ (i.e., each user is fully altruistic and cares enough about other users), the SGUM game becomes the network utility maximization problem, which aims to maximize the system-wide utility. The SGUM framework in this study is applicable to  general social graphs and hence can bridge the gap between non-cooperative game and network utility maximization -- two traditionally disjoint paradigms for network optimization. Roughly speaking, we can interpret the SGUM framework from the perspective of information sharing. If in the system more users would like to share the information of their individual utilities, then more informed decision making for optimizing the collective performance can be achieved,  and hence the system performance could be improved as the social link density increases.

The SGUM is a general framework that can be applied for many networking applications. To get a more concrete sense of the framework, in the following sections, we will study its applications in database assisted spectrum access, power control, and random access control.

\section{\label{sec:System-Model}Social Group Utility Maximization For Database Assisted Spectrum Access}

In this section we apply the SGUM framework for the database assisted spectrum access.


\subsection{SGUM Game Formulation}

\begin{figure}[tt]
\begin{center}
\includegraphics[scale=0.3]{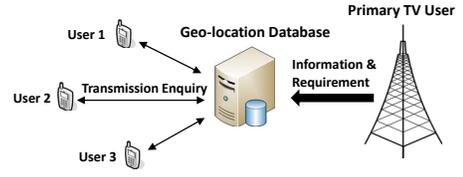}
\caption{\label{fig:Database}An illustration of database assisted spectrum access}
\end{center}
\end{figure}

We consider a set of white-space users $\mathcal{N}=\{1,2,...,N\}$
where $N$ is the total number of users. We denote the set of TV
channels as $\mathcal{M}=\{1,2,...,M\}$. According to the recent ruling by FCC\cite{FCC}, to protect the incumbent primary TV users, each white-space user $n\in\mathcal{N}$ will first
send a spectrum access request message containing its geo-location
information to a Geo-location database (see Figure \ref{fig:Database} for an illustration). The database then feeds
back the set of vacant channels $\mathcal{M}_{n}\in\mathcal{M}$ and
the allowable transmission power level $P_{n}$ to user $n$. The ruling by FCC indicates that the allowable transmission
power limit for personal/portable white-space devices (e.g., mobile
phones) is $100$ mW \cite{FCC}. For ease of exposition, we hence assume
that each user $n$ accesses the white-space spectrum with the same
power level. Each user $n$ then chooses a feasible channel $a_{n}$ from the
vacant channel set $\mathcal{M}_{n}$ for data transmission.  Although the database-assisted approach obviates the need of spectrum sensing by individual users, it remains challenging to achieve reliable distributed spectrum access, because many different white-space users may choose to access the same vacant channel and thus incur severe interference to each other \cite{yangton13,Database2013}.

To stimulate effective cooperation among users for interference mitigation, we leverage the social ties among users and apply the SGUM approach. To capture the physical coupling and account for the accumulative nature of interference, we construct the physical interference graph $\mathcal{G}^{p}=\{\mathcal{N},\mathcal{E}^{p}\}$ based on the physical interference model \cite{gupta2000capacity}. Here the set of white-space users $\mathcal{N}$
is the vertex set, and $\mathcal{E}^{p}\triangleq\{(n,m):e_{nm}^{p}=1,\forall n,m\in\mathcal{N}\}$
is the edge set where $e_{nm}^{p}=1$ if and only if users $n$
and $m$ can receive interference from each other\footnote{The
physical interference model enables us to capture the cumulative nature of interference by defining
the individual utility function  based on user's total received interference from other users. This is different from the
protocol interference graph model where two neighboring users' data transmissions are blocked once they transmit
simultaneously over the same vacant channel.}. To account for the potential interference from every user choosing the same vacant channel, we can regard the interference graph $\mathcal{G}^{p}$ as a complete graph. Nevertheless, our model can also apply for the case that the underlying interference graph is undirected but not complete. This can be applicable, for example, when we would like to factor those users can generate significant impact only and neglect the users that are too far away (e.g., with the interference power even less than the noise) \cite{zhou2013practical}. Let $\boldsymbol{a}=(a_{1},...,a_{N})\in\prod_{n=1}^{N}\mathcal{M}_{n}$
be the channel selection profile of all users. Given the channel
selection profile $\boldsymbol{a}$, the total interference received by
user $n$ can be computed as
\begin{equation}
\gamma_{n}(\boldsymbol{a})=\sum_{m\in\mathcal{N}_{n}^{p}}P_{m}d_{mn}^{-\alpha}I_{\{a_{n}=a_{m}\}}+\omega_{a_{n}}^{n}.\label{eq:interference1}
\end{equation}
Here $\alpha$ is the path loss factor and $I_{\{A\}}$ is an indicator
function with $I_{\{A\}}=1$ if the event $A$ is true and $I_{\{A\}}=0$
otherwise. Furthermore, $\omega_{a_{n}}^{n}$ denotes the noise power
including the interference from primary TV users on the channel $a_{n}$. We then define the individual utility function $U_{n}(\boldsymbol{a})$
as
\begin{align}
U_{n}(\boldsymbol{a})  = -\gamma_{n}(\boldsymbol{a}) = -\sum_{m\in\mathcal{N}_{n}^{p}}P_{m}d_{mn}^{-\alpha}I_{\{a_{n}=a_{m}\}}-\omega_{a_{n}}^{n}.\label{eq:utility}
\end{align}
Here the negative sign comes from the convention that utility functions
are typically the ones to be maximized. The individual utility of user
$n$ reflects the fact that each user $n$ has interest to reduce
its own received interference.  Similar to many previous studies such as \cite{ramachandran2006interference,tutschku1999interference,chen2011cross}, we focus on the objective of interference minimization. By exploiting the additive structure of the accumulative interference functions, we are going to analyze the SGUM game for interference minimization by resorting to the useful tool of potential game \cite{monderer1996potential}. Note that, although from the individual utility perspective maximizing a user's data rate is equivalent to minimizing its received interference, this may not be true from the social group utility perspective. Due to the complicated couplings among users in both social and physical domains, the SGUM game of data rate maximization is technically very challenging, and the potential game approach can not apply in this case.

Next, given the social ties (i.e., $w_{ij}$) among the users on the social graph $\mathcal{G}^{s}$, we can define the social group utility function $S_i$  of user $i$ according to (\ref{eq:SocialUtility0}). We then formulate the database assisted spectrum access problem with social ties as a SGUM game $\Gamma=(\mathcal{N},\{a_{n}\},\{S_{n}\})$,
where the set of white-space users $\mathcal{N}$ is the set of players,
the set of vacant channels $\mathcal{M}_{n}$ is the set of strategies
for each player $n$, and the social group utility function
$S_{n}$ of each user $n$ is the payoff function of player $n$.

\subsection{Properties of SGUM game}

We next study the existence of SNE of the SGUM game for database assisted spectrum access. Here we resort
to the tool of potential game \cite{monderer1996potential}.
\begin{defn}
A game is called a potential game if it admits a potential function
$\Phi(\boldsymbol{a})$ such that for every $n\in\mathcal{N}$ and
$a_{-n}\in\prod_{i\neq n}\mathcal{M}_{i}$, for any $a_{n},a_{n}^{'}\in\mathcal{M}_{n}$,
\begin{align}
S_{n}(a_{n}^{'},a_{-n})-S_{n}(a_{n},a_{-n})=\Phi(a_{n}^{'},a_{-n})-\Phi(a_{n},a_{-n}). \label{eq:Potential}
\end{align}
\end{defn}

An appealing property of a potential game with a finite strategy set is that it always admits
a Nash equilibrium, and any strategy profile that maximizes the potential
function $\Phi(\boldsymbol{a})$ is a Nash equilibrium \cite{monderer1996potential}. For the SGUM game  $\Gamma$ for database assisted spectrum access, we can show that it is a potential
game. For ease of exposition, we first introduce the \emph{physical-social
graph} $\mathcal{G}^{sp}=\{\mathcal{N},\mathcal{E}^{sp}\}$ to capture both physical coupling and social coupling simultaneously. Here the
vertex set is the same as the user set $\mathcal{N}$ and the
edge set is given as $\mathcal{E}^{sp}=\{(n,m):e_{nm}^{sp}\triangleq e_{nm}^{s}\cdot e_{nm}^{p}=1,\forall n,m\in\mathcal{N}\}$
where $e_{nm}^{sp}=1$ if and only if users $n$ and $m$ have
social tie between each other (i.e., $e_{nm}^{s}=1$) and can also
generate interference to each other (i.e., $e_{nm}^{p}=1$). We denote
the set of users that have social ties and can also generate
interference to user $n$ as $\mathcal{N}_{n}^{sp}=\{m:e_{nm}^{sp}=1,\forall m\in\mathcal{N}\}$.

Based on the physical-social graph $\mathcal{G}^{sp}$, we show in
Theorem \ref{thm:The-social-physical-game} that the SGUM game $\Gamma$
is a potential game with the following potential function
\begin{align}
\Phi(\boldsymbol{a})= & \underbrace{-\frac{1}{2}\sum_{n=1}^{N}\sum_{m\in\mathcal{N}_{n}^{p}}P_{m}d_{mn}^{-\alpha}I_{\{a_{n}=a_{m}\}} -\sum_{n=1}^{N}\omega_{a_{n}}^{n}}_{\Phi_{1}(\boldsymbol{a}):\mbox{ due to Physical Coupling}} \nonumber \\
 & \underbrace{-\frac{1}{2}\sum_{n=1}^{N}\sum_{m\in\mathcal{N}_{n}^{sp}}w_{nm}P_{m}d_{mn}^{-\alpha}I_{\{a_{n}=a_{m}\}}}_{\Phi_{2}(\boldsymbol{a}):\mbox{ due to Social Coupling}}.\label{eq:potential2}
\end{align}

The potential function in (\ref{eq:potential2}) can be decomposed into two parts:
$\Phi_{1}(\boldsymbol{a})$ and $\Phi_{2}(\boldsymbol{a})$. The first part $\Phi_{1}(\boldsymbol{a})$ reflects the system-wide interference level (including background noise)
due to physical coupling in the physical domain and the second part $\Phi_{2}(\boldsymbol{a})$ captures the interdependence of user utilities (i.e., the received interferences)  due to social
coupling in the social domain.

\begin{thm}
\label{thm:The-social-physical-game}When the transmission power of the users are the same (i.e., $P_n=P_m$), the physical interference graph is undirected and the social tie between two users is symmetric (i.e., $w_{nm}=w_{mn}$), the SGUM game for database assisted spectrum access is a potential game
with the potential function \textup{$\Phi(\boldsymbol{a})$} given
in (\ref{eq:potential2}), and hence has a SNE.\end{thm}

The proof can be found in the online appendix \cite{SGUM2016}. Note that when $w_{nm}=0$ for any user
$n,m\in\mathcal{N}$ (i.e., all users are selfish), the potential function $\Phi(\boldsymbol{a})  =  \Phi_{1}(\boldsymbol{a})$, which does not involve the social coupling part $\Phi_{2}(\boldsymbol{a})$. In this case, the SGUM game $\Gamma$ for database assisted spectrum access degenerates to the non-cooperative spectrum access game. When $w_{nm}=1$ for any user $n,m\in\mathcal{N}$ (i.e., all users are fully altruistic), the potential function $\Phi(\boldsymbol{a})  =  \sum_{n=1}^{N}U_n(\boldsymbol{a})$,  which is the system-wide utility.  In this case, the SGUM becomes the network utility maximization.

We next design a distributed spectrum access algorithm that
can achieve the SNE of the SGUM game $\Gamma$ for database assisted spectrum access.

\subsection{\label{sec:algorithm}Distributed Spectrum Access Algorithm For Social Group Utility Maximization}

In this section we study the distributed spectrum access algorithm
design.

\subsubsection{Algorithm Design Principles}
According to the property of potential game, any channel selection
profile $\boldsymbol{a}$ that maximizes the potential function $\Phi(\boldsymbol{a})$
is a Nash equilibrium \cite{monderer1996potential}. We hence design a distributed spectrum
access algorithm that achieves the SNE of the SGUM $\Gamma$ by maximizing the potential function $\Phi(\boldsymbol{a})$. Note that since the SGUM game is a potential game, one can choose to use the asynchronous best response update approach such that in each iteration a user is selected to optimize its own strategy \cite{nisan2007algorithmic}. However, there are several drawbacks associated with  the asynchronous best response update approach for our problem: (1) the asynchronous best response update is sensitive to the initial state as well as the update sequence. This implies that given different initial channel selections and different channel update schedules, the spectrum access algorithm based on the asynchronous best response update may reach different convergent points; (2) as the potential function of the SGUM game captures both physical and social couplings among users for interference minimization, a larger value of potential function  which implies a lower interference level is hence more desirable. However, the asynchronous best response update (such that each user locally and greedily optimizes its own strategy) may get stuck at a local optimum of the potential function. Thus, this motivates us to design the distributed spectrum access algorithm based on Markov chain design,  which can converge to a unique stationary distribution regardless of the given initial channel selection and update sequence. Moreover, we show that the algorithm can asymptotically converge to the global optimum of the potential function with an arbitrarily large probability. Note that one may consider to adopt the learning algorithm in \cite{pradelski2012learning} to achieve an efficient equilibrium. However, the algorithm in \cite{pradelski2012learning} requires the strong assumptions of synchronization and interdependence, which are difficult to be satisfied in our case.

To proceed, first consider the problem that the  users collectively
compute the optimal channel selection profile such that the potential
function is maximized, i.e.,
\begin{equation}
\max_{\boldsymbol{a}\in\Omega\triangleq\prod_{n=1}^{N}\mathcal{M}_{n}}\Phi(\boldsymbol{a}).\label{eq:pp3}
\end{equation}
The problem (\ref{eq:pp3}) involves  a combinatorial optimization over the discrete
solution space $\Omega$. In general, it is very challenging
to solve such a problem exactly especially when the system size is large (i.e., the
solution space $\Omega$ is large).

We then consider to approach the potential function maximization solution approximately.
To proceed, we first write the problem (\ref{eq:pp3}) as the following
equivalent randomized problem:
\begin{eqnarray}
\max_{(q_{\boldsymbol{a}}\geq0:\boldsymbol{a}\in\Omega)} & \sum_{\boldsymbol{a}\in\Omega}q_{\boldsymbol{a}}\Phi(\boldsymbol{a})\label{eq:pp4}\\
\mbox{s.t.} & \sum_{\boldsymbol{a}\in\Omega}q_{\boldsymbol{a}}=1,\nonumber
\end{eqnarray}
where $q_{\boldsymbol{a}}$ is the probability that channel selection
profile $\boldsymbol{a}$ is adopted. Obviously, the optimal solution
to problem (\ref{eq:pp4}) is to choose the optimal channel selection
profiles with probability one. It is known from the Markov approximation approach \cite{chen2010markov} that problem
(\ref{eq:pp4}) can be approximated by the following convex optimization
problem:
\begin{eqnarray}
\max_{(q_{\boldsymbol{a}}\geq0:\boldsymbol{a}\in\Omega)} & \sum_{\boldsymbol{a}\in\Omega}q_{\boldsymbol{a}}\Phi(\boldsymbol{a})-\frac{1}{\theta}\sum_{\boldsymbol{a}\in\Omega}q_{\boldsymbol{a}}\ln q_{\boldsymbol{a}}\label{eq:pp5}\\
\mbox{s.t.} & \sum_{\boldsymbol{a}\in\Omega}q_{\boldsymbol{a}}=1,\nonumber
\end{eqnarray}
where $\theta$ is the parameter that controls the approximation ratio. Note that the approximation in (\ref{eq:pp5}) can guarantee the asymptotic optimality. This is because that when $\theta\rightarrow\infty$, the problem (\ref{eq:pp5})
boils down to  exactly the same as problem (\ref{eq:pp4}). That is, when
$\theta\rightarrow\infty$, the optimal solutions that maximize the
potential function $\Phi(\boldsymbol{a})$ will be selected with probability
one. Moreover, the approximation in (\ref{eq:pp5}) enables us to obtain the close-form solution, which facilitates the distributed
algorithm design later. More specifically, by the KKT conditions \cite{boyd2004convex},
the optimal solution to problem (\ref{eq:pp5}) is given as
\begin{equation}
q_{\boldsymbol{a}}^{*}=\frac{\exp(\theta\Phi(\boldsymbol{a}))}{\sum_{\hat{\boldsymbol{a}}\in\Omega}\exp(\theta\Phi(\hat{\boldsymbol{a}}))}.\label{eq:qq1}
\end{equation}

Inspired by the idea of Glauber dynamics \cite{rajagopalan2008distributed,jiang2010distributed,ni2012q,hegde2012simulation},  based on (\ref{eq:qq1}) we then design a self-organizing algorithm
such that the asynchronous channel selection updates of the  users
form a Markov chain (with the system state as the channel selection
profile $\boldsymbol{a}$ of all users). As long as the Markov chain
converges to the stationary distribution as given in (\ref{eq:qq1}),
we can approach the Nash equilibrium channel selection profile that
maximizes the potential function by setting a large enough parameter
$\theta$. We should emphasize that the existing works such as \cite{rajagopalan2008distributed,jiang2010distributed,ni2012q,hegde2012simulation} focus on the NUM problem, whereas in this paper we consider the SGUM problem, which has significant differences. For example,  for NUM, the optimization objective
of each user is aligned. In contrast, for SGUM, different users have different optimization objectives due to the heterogenous social coupling among them. The existing works use the transition probability matrix for Markov chain design based on the canonical Glauber dynamics.  In contrast, in this paper we leverage the structural property of the SGUM game to devise the transition probability for the spectrum access Markov chain. We also quantify the impact of social tie on the networking performance, which has not been found in
the existing works.

\subsubsection{Markov Chain Design for Distributed Spectrum Access}

\begin{algorithm}[tt]
\begin{algorithmic}[1]
\State \textbf{initialization:}
\State \hspace{0.4cm} \textbf{set} the parameter $\theta$ and the channel update rate $\tau_{n}$.
\State \hspace{0.4cm} \textbf{choose} a channel $a_{n}\in\mathcal{M}_{n}$ randomly for each  user $n\in\mathcal{N}.$
\State \textbf{end initialization\newline}

\Loop{ for each user $n\in\mathcal{N}$ in parallel:}
        \State \textbf{compute} the social group utility  $S_{n}(a_{n},a_{-n})$ on the chosen channel $a_{n}$.
        \State \textbf{generate}  a timer value following  the exponential distribution with the mean equal to $\frac{1}{\tau_{n}}$.
        \State \textbf{count down} until the timer expires.
        \If{the timer expires}
            \State \textbf{choose}  a new channel $a_{n}^{'}\in\mathcal{M}_{n}$ randomly.
            \State \textbf{compute} the social group utility $S_{n}(a_{n}^{'},a_{-n})$ on the new channel $a_{n}^{'}$.
            \State \textbf{stay in} the new channel $a_{n}^{'}$ with probability $\frac{\exp\left(\theta S_{n}(a_{n}^{'},a_{-n})\right)}{\max\{\exp\left(\theta S_{n}(a_{n}^{'},a_{-n})\right), \exp\left(\theta S_{n}(a_{n},a_{-n})\right)\}}$, Or \textbf{move back} to the original channel $a_{n}$ with probability $1-\frac{\exp\left(\theta S_{n}(a_{n}^{'},a_{-n})\right)}{\max\{\exp\left(\theta S_{n}(a_{n}^{'},a_{-n})\right), \exp\left(\theta S_{n}(a_{n},a_{-n})\right)\}}$.
        \EndIf
\EndLoop
\end{algorithmic}
\caption{\label{alg:DSSA} Distributed Spectrum Access Algorithm For Social Group Utility Maximization}
\end{algorithm}

Motivated by the adaptive CSMA mechanism \cite{jiang2010distributed,chen2010markov}, we propose
a distributed spectrum access algorithm in Algorithm \ref{alg:DSSA} such that each  user
$n$ updates its channel selection according to a timer value that
follows the exponential distribution with a rate of $\tau_{n}$.

Appealing to the property of exponential distributions, we have that the probability that more than one users generate the same timer value and update their channels simultaneously equals zero. Since one  user will activate for the channel selection update at
a time, the direct transitions between two system states $\boldsymbol{a}$
and $\boldsymbol{a}^{'}$ are feasible if these two system states
differ by one and only one  user's channel selection. We also denote
the set of system states that can be transited directly from the state
$\boldsymbol{a}$ as $\Delta_{\boldsymbol{a}}=\{\boldsymbol{a}^{'}\in\Omega:|\{\boldsymbol{a}^{'}\cup\boldsymbol{a}\}\backslash\{\boldsymbol{a}^{'}\cap\boldsymbol{a}\}|=2\}$,
where $|\cdot|$ denotes the size of a set. According to (\ref{eq:SocialUtility0}),
a  user $n$ can compute the social group utility $S_{n}(\boldsymbol{a})$ by locally enquiring
the users having social ties with it about their received interferences.
Then user $n$ will randomly choose a new channel $a_{n}^{'}\in\mathcal{M}_{n}$
and stay in this channel with a probability of
\begin{equation}
\frac{\exp\left(\theta S_{n}(a_{n}^{'},a_{-n})\right)}{\max\{\exp\left(\theta S_{n}(a_{n}^{'},a_{-n})\right), \exp\left(\theta S_{n}(a_{n},a_{-n})\right)\}}. \label{eq:kkkkk}
\end{equation}
The underlying rationale  behind (\ref{eq:kkkkk}) is as follows.
When $S_{n}(a_{n}^{'},a_{-n})\geq S_{n}(a_{n},a_{-n})$ (i.e., the new channel $a_{n}^{'}$ offers the better performance), user $n$ will stay in the new channel $a_{n}^{'}$ with probability one. According to the property of potential game in (\ref{eq:Potential}), we know that choosing the new channel $a_{n}^{'}$ can help to increase both user $n$'s social group utility  $S_{n}(\boldsymbol{a})$ and the potential function $\Phi(\boldsymbol{a})$ of the SGUM game. When $S_{n}(a_{n}^{'},a_{-n}) < S_{n}(a_{n},a_{-n})$ (i.e., the original channel $a_{n}$ offers the better performance), user $n$ will switch back to the original channel $a_{n}$ with a probability of $1-\frac{\exp\left(\theta S_{n}(a_{n}^{'},a_{-n})\right)}{\exp\left(\theta S_{n}(a_{n},a_{-n})\right)}$. That is, the probability that user $n$ will switch back to the original channel $a_{n}$ increases with the performance gap $S_{n}(a_{n},a_{-n})-S_{n}(a_{n}^{'},a_{-n})$. We would like to emphasize that such probabilistic channel selections are necessary such that all the users can explore the feasible channel selection space to prevent the algorithm from getting stuck at a local optimum.

Then from a system-wide perspective, the probability of transition from state $(a_{n},a_{-n})$ to
$(a_{n}^{'},a_{-n})$ due to user $n$'s channel selection update is given as
\begin{equation}
\frac{1}{|\mathcal{M}_{n}|}\frac{\exp\left(\theta S_{n}(a_{n}^{'},a_{-n})\right)}{\max\{\exp\left(\theta S_{n}(a_{n}^{'},a_{-n})\right), \exp\left(\theta S_{n}(a_{n},a_{-n})\right)\}}.\label{eq:ppppp}
\end{equation}
Since each user $n$ activates
its channel selection update according to the countdown timer mechanism
with a rate of $\tau_{n}$, hence if $\boldsymbol{a}^{'}\in\Delta_{\boldsymbol{a}}$,
the transition rate from state $\boldsymbol{a}$ to state $\boldsymbol{a}^{'}$
is given as
\begin{equation}
q_{\boldsymbol{a},\boldsymbol{a}^{'}}=\frac{\tau_{n}}{|\mathcal{M}_{n}|}\frac{\exp\left(\theta S_{n}(a_{n}^{'},a_{-n})\right)}{\max\{\exp\left(\theta S_{n}(a_{n}^{'},a_{-n})\right), \exp\left(\theta S_{n}(a_{n},a_{-n})\right)\}}.\label{eq:PPP4}
\end{equation}
Otherwise, we have $q_{\boldsymbol{a},\boldsymbol{a}^{'}}=0$. We
show in Theorem \ref{thm:The-distributed-D2D} that the spectrum access
Markov chain is time reversible. Time reversibility means that when
tracing the Markov chain backwards, the stochastic behavior of the
reverse Markov chain remains the same. A nice property of a time reversible
Markov chain is that it always admits a unique stationary distribution,
which is independent of the initial system state. This implies that
given any initial channel selections the distributed spectrum
access algorithm can drive the system converging to the stationary
distribution given in (\ref{eq:qq1}).
\begin{thm}
\label{thm:The-distributed-D2D}The distributed spectrum access algorithm
induces a time-reversible Markov chain with the unique stationary
distribution as given in (\ref{eq:qq1}).\end{thm}

The proof is given in the online appendix \cite{SGUM2016}.  To implement the distributed spectrum access algorithm, as mentioned in Section \ref{sec:framework}, each user can first identify the social tie with other users prior to the spectrum access process. Then a user can locally measure its received interference and exchange this information with its social friends for computing the social group utility, which will be used for channel selection update.

We next analyze the convergence time of the proposed algorithm in terms
of mixing time in Markov chain \cite{levin2009markov}. Let $\boldsymbol{q}^{*}$ denote the
stationary distribution in (\ref{eq:qq1}) and $\boldsymbol{q}_{t}(\boldsymbol{a}_{0})$
denote the distribution of the spectrum access Markov chain at time
$t$ for a given initial channel selection state $\boldsymbol{a}_{0}$.
Then, according to \cite{levin2009markov}, for $\epsilon>0$, the mixing time of
the chain is defined by
\[
T_{\epsilon}\triangleq\inf\{t\geq0:\max_{\boldsymbol{a}_{0}\in\Omega}||\boldsymbol{q}_{t}(\boldsymbol{a}_{0})-\boldsymbol{q}^{*}||_{TV}\leq\epsilon\},
\]
where the total variation distance  is given by  \[||\boldsymbol{q}_{t}(\boldsymbol{a}_{0})-\boldsymbol{q}^{*}||_{TV}\triangleq\frac{1}{2}\sum_{\boldsymbol{a}\in\Omega}|q_{\boldsymbol{a},t}(\boldsymbol{a}_{0}),q_{a}^{*}|.\]
We further denote $M_{\max}=\max_{n\in\mathcal{N}}|\mathcal{M}_{n}|$,
$M_{min}=\min_{n\in\mathcal{N}}|\mathcal{M}_{n}|>0$, $\tau_{\max}=\max_{n\in\mathcal{N}}\tau_{n}$,
$\tau_{\min}=\min_{n\in\mathcal{N}}\tau_{n},$ $\Phi_{\max}=\max_{\boldsymbol{a}\in\Omega}\Phi(\boldsymbol{a}),$
$\Phi_{\min}=\min_{\boldsymbol{a}\in\Omega}\Phi(\boldsymbol{a})$.
By resorting to the tools of spectral analysis and path coupling \cite{chen2010markov,levin2009markov}, we can show in Theorem \ref{thm:mixing} the convergence time of the spectrum
access Markov chain.
\begin{thm}\label{thm:mixing}
For a general $\theta\in(0,\infty)$, the mixing time of the spectrum
access Markov chain is upper-bounded as
\begin{eqnarray*}
T_{\epsilon} & \leq & \frac{NM_{\max}^{2N+3}\tau_{\max}}{M_{\min}\tau_{\min}^{2}}\exp\left(4\theta(\Phi_{\max}-\Phi_{\min})\right)\\
 &  & \times\left[2\ln\frac{1}{2\epsilon}+N\ln M_{\max}+\theta(\Phi_{\max}-\Phi_{\min})\right].
\end{eqnarray*}
When $\theta\in(0,\theta_{th})$ where $\theta_{th}\triangleq\frac{1}{\Phi_{\max}-\Phi_{\min}}\ln\left(\frac{NM_{\min}^{2}\tau_{\min}}{(N-1)M_{\max}\tau_{\max}}\right)$,
we have a tighter upper bound as
\begin{align*}
T_{\epsilon} \leq \ln\frac{N}{\epsilon}\frac{M_{\min}}{M_{\max}\tau_{\max}}\frac{\exp(\theta(\Phi_{\max}-\Phi_{\min}))}{\frac{M_{\min}^{2}\tau_{\min}}{M_{\max}\tau_{\max}}+\frac{1-N}{N}\exp(\theta(\Phi_{\max}-\Phi_{\min}))}.
\end{align*}
\end{thm}

The proof is given in the online appendix \cite{SGUM2016}. Theorem \ref{thm:mixing} provides an upper-bound of the convergence time of the distributed spectrum access algorithm. In particular, it reveals when the parameter  $\theta$ is small, the proposed spectrum access algorithm scales very well (in the order of $\mathcal{O}(\ln N)$) with the increase of user size $N$.

Note that compared with existing studies \cite{rajagopalan2008distributed,jiang2010distributed,ni2012q,hegde2012simulation} on applying the Glauber dynamics, we have the following new contributions for the convergence time analysis: 1) we use the spectral gap analysis to show the mixing time upper-bound for a general parameter $\theta$, and the existing studies only show the mixing time upper-bound when the parameter $\theta$ is within some specific range by adopting the path coupling method. This is due to the restriction that the path coupling approach imposes some structural condition on the Markov chain; 2) since in our problem the
underlying optimization objective and the Markov chain structure are different
from existing studies, we have carefully constructed a discrete-time Markov chain by uniformization of the original spectrum access Markov chain, and showed that when $\theta<\theta_{th}$ the constructed Markov chain satisfies the one-step path coupling condition. This leads to significant differences from the existing works when applying the path coupling method.


\subsubsection{Performance Analysis}

According to Theorem \ref{thm:The-distributed-D2D}, we can achieve
the SNE that maximizes the potential
function $\Phi(\boldsymbol{a})$ of the SGUM game $\Gamma$
by setting $\theta\rightarrow\infty$. However, in practice one has to choose to implement
only a finite value of $\theta$ (e.g., to reduce the convergence time). Let $\Phi_{\theta}=\sum_{a\in\Omega}q_{a}^{*}\Phi(\boldsymbol{a})$
be the expected potential by Algorithm \ref{alg:DSSA} with a given $\theta$ and $\Phi^{*}=\max_{a\in\Omega}\Phi(\boldsymbol{a})$
be the maximum potential. We show in Theorem \ref{thm:For-the-distributed-1}
that, when a large enough $\theta$ is adopted in practice, the
gap between $\Phi_{\theta}$ and $\Phi^{*}$ is negligible. Combining Theorems \ref{thm:mixing} and \ref{thm:For-the-distributed-1}, we see that in practice by tuning the parameter $\theta$, we can significantly reduce the convergence time (at the order of $e^{\theta}$) at the cost of a relatively small performance loss (at the order of $\theta^{-1}$). This has also been corroborated  by the numerical result in Section \ref{sec:simulation}.

\begin{thm}
\label{thm:For-the-distributed-1}For the distributed spectrum access
algorithm with a finite $\theta$, we have that $0\leq\Phi^{*}-\Phi_{\theta}\leq\frac{1}{\theta}\sum_{n=1}^{N}\ln|\mathcal{M}_{n}|,$
where $|\mathcal{M}_{n}|$ denotes the number of vacant channels of  user $n$.\end{thm}

The proof is given in the online appendix \cite{SGUM2016}. We next benchmark the performance of the solution by the distributed spectrum access algorithm with respect to the system optimal solution. Let $V(\boldsymbol{a})$ be the total individual utility received by all the users under the
channel selection profile $\boldsymbol{a}$ (i.e., $V(\boldsymbol{a})=\sum_{n=1}^{N}U_{n}(\boldsymbol{a})$) and $V_{\theta}$ be the total utility of the solution by the distributed spectrum access algorithm with a finite $\theta$ (i.e., $V_{\theta}=\sum_{\boldsymbol{a}\in\Omega}q_{\boldsymbol{a}}^{*}V(\boldsymbol{a})$).
We denote $\overline{\boldsymbol{a}}$ as the NUM solution
that maximizes the system-wide utility (i.e., $\overline{\boldsymbol{a}}=\arg\max_{a\in\Omega}V(\boldsymbol{a})$).
We then define the performance gap $\rho_{\theta}$ as the difference between the total utility received at the NUM solution
$\overline{\boldsymbol{a}}$ and that of the distributed spectrum access algorithm with a finite $\theta$ (i.e., $\rho_{\theta}=V(\overline{\boldsymbol{a}})-V_{\theta}$).
We have the following result.
\begin{thm}
\label{thm:The-performance-gap1}For a finite $\theta < \infty$,  the performance gap $\rho_{\theta}$ of the distributed spectrum access algorithm is at most
\begin{align*}
 & \frac{1}{\theta}\sum_{n=1}^{N}\ln|\mathcal{M}_{n}| + \frac{1}{2}\sum_{n=1}^{N}\sum_{m\in\mathcal{N}_{n}^{sp}}(1-w_{nm})P_{m}d_{mn}^{-\alpha} \\
 & +\frac{1}{2}\sum_{n=1}^{N}\sum_{m\in\mathcal{N}_{n}^{p}\backslash\mathcal{N}_{n}^{sp}}P_{m}d_{mn}^{-\alpha}.
\end{align*}
\end{thm}

The proof is given in the online appendix \cite{SGUM2016}. Let $\hat{\boldsymbol{a}}$ as the convergent SNE by
the distributed spectrum access algorithm when $\theta \rightarrow \infty$ (i.e., $\hat{\boldsymbol{a}}=\arg\max_{a\in\Omega}\Phi(\boldsymbol{a})$). According to Theorem \ref{thm:The-performance-gap1}, the performance gap of the SNE $\hat{\boldsymbol{a}}$ is given as follows.

\begin{cor}
\label{thm:The-performance-gap} When $\theta \rightarrow \infty$, the performance gap of the SNE $\hat{\boldsymbol{a}}$ by the distributed spectrum access algorithm is at most
\begin{align*}
\frac{1}{2}\sum_{n=1}^{N}\sum_{m\in\mathcal{N}_{n}^{sp}}(1-w_{nm})P_{m}d_{mn}^{-\alpha} +\frac{1}{2}\sum_{n=1}^{N}\sum_{m\in\mathcal{N}_{n}^{p}\backslash\mathcal{N}_{n}^{sp}}P_{m}d_{mn}^{-\alpha}.
\end{align*}
\end{cor}

From the results above we see that the upper-bound
of the performance gap decreases as the strength of social
tie $w_{nm}$ among users increases.  When $w_{nm}=1$ for any user $n,m\in\mathcal{N}$ (i.e., all users
are fully altruistic), the SGUM becomes
the NUM and the performance gap is zero.  In Section \ref{sec:simulation}, we also evaluate the performance of the SGUM solution by real social data traces. Numerical results demonstrate that the performance gap between the SGUM solution and the NUM solution is at most $15\%$.

\subsection{\label{sec:simulation}Numerical Evaluation}

In this section, we evaluate the SGUM solution for database assisted spectrum access by numerical studies based on both Erdos-Renyi social graphs and real trace based social graphs.

\subsubsection{Social Graph with 8 White-Space Users}
\begin{figure}
\centering
\includegraphics[scale=0.6]{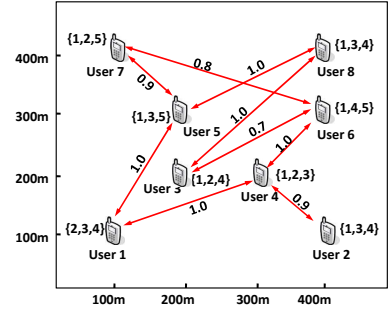}
\caption{\label{fig:A-square-area}A square area of a length of 500 m with 8
scattered white-space users}
\end{figure}

\begin{figure*}[tt]
\begin{minipage}[t]{0.32\linewidth}
\centering
\includegraphics[scale=0.35]{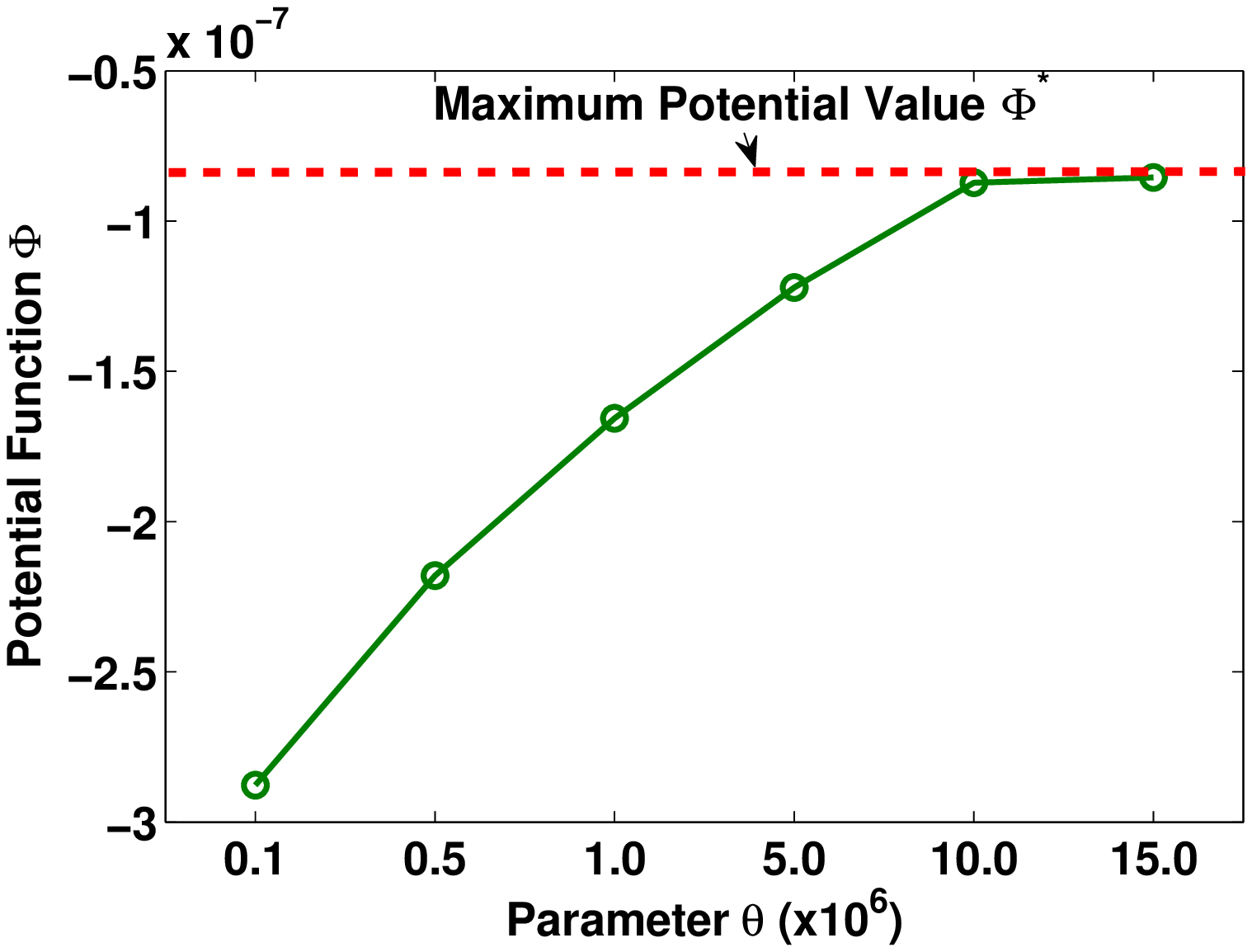}
\caption{\label{fig:Potential-value-}The convergent potential value $\Phi$
with different parameters $\theta$}
\end{minipage}
\hfill
\begin{minipage}[t]{0.32\linewidth}
\centering
\includegraphics[scale=0.35]{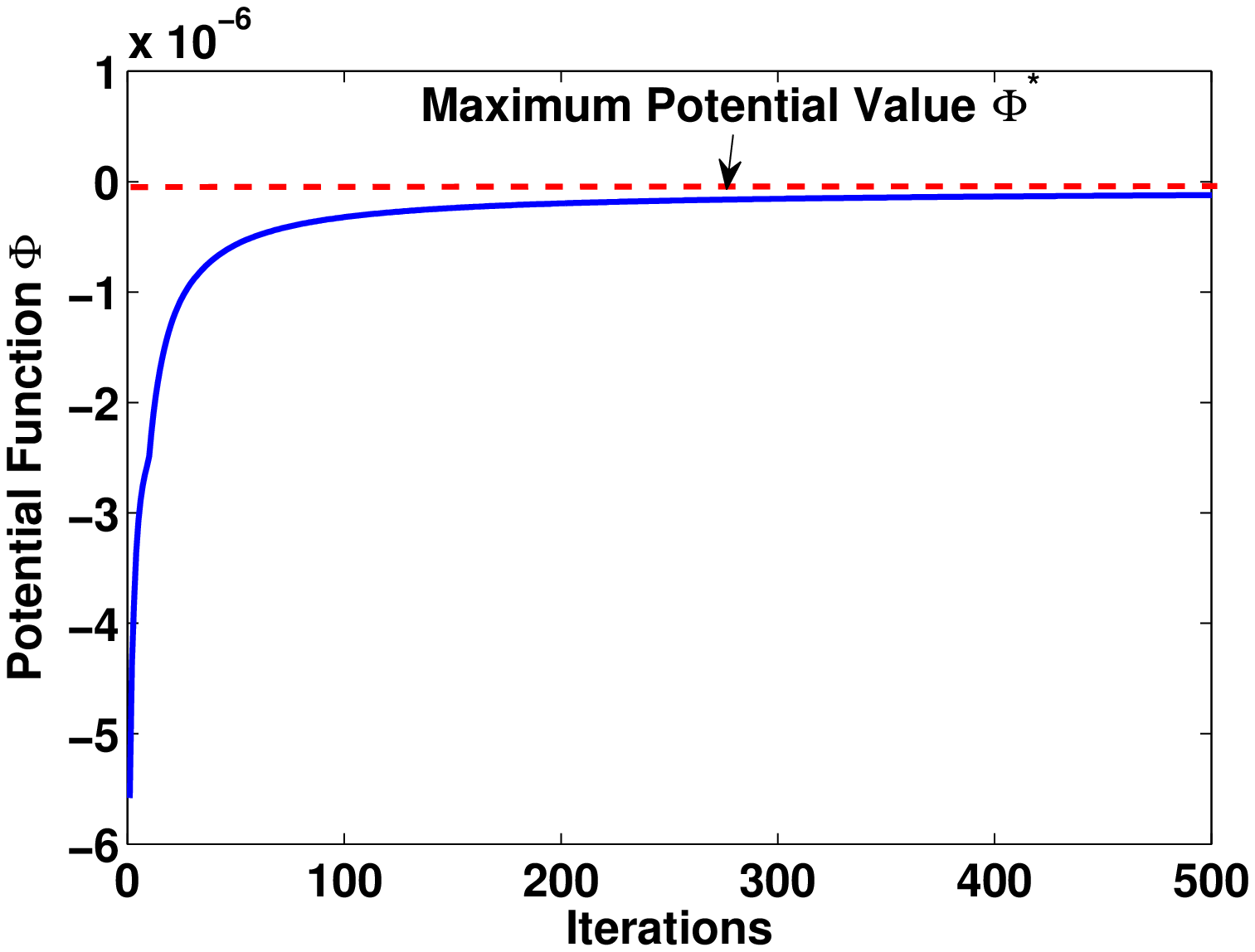}
\caption{\label{fig:Dynamics-of-potential}Dynamics of potential value $\Phi$
when $\theta=10^{6}$}
\end{minipage}
\hfill
\begin{minipage}[t]{0.32\linewidth}
\centering
\includegraphics[scale=0.342]{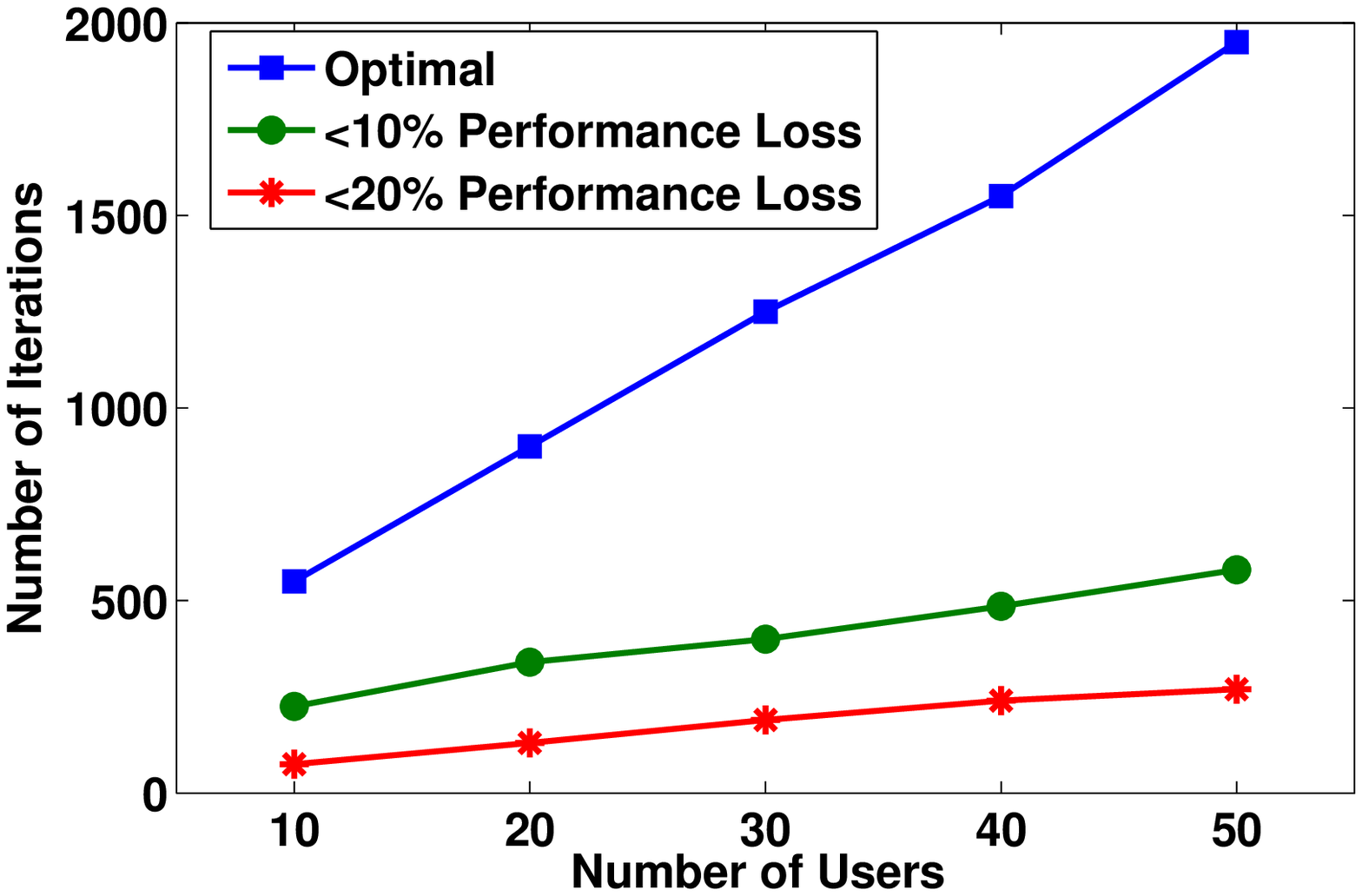}
\caption{\label{fig:iteration}Number of iterations with different maximum allowable performance loss}
\end{minipage}
\end{figure*}

%
%
%
%

\begin{figure*}[tt]
\begin{minipage}[t]{0.32\linewidth}
\centering
\includegraphics[scale=0.35]{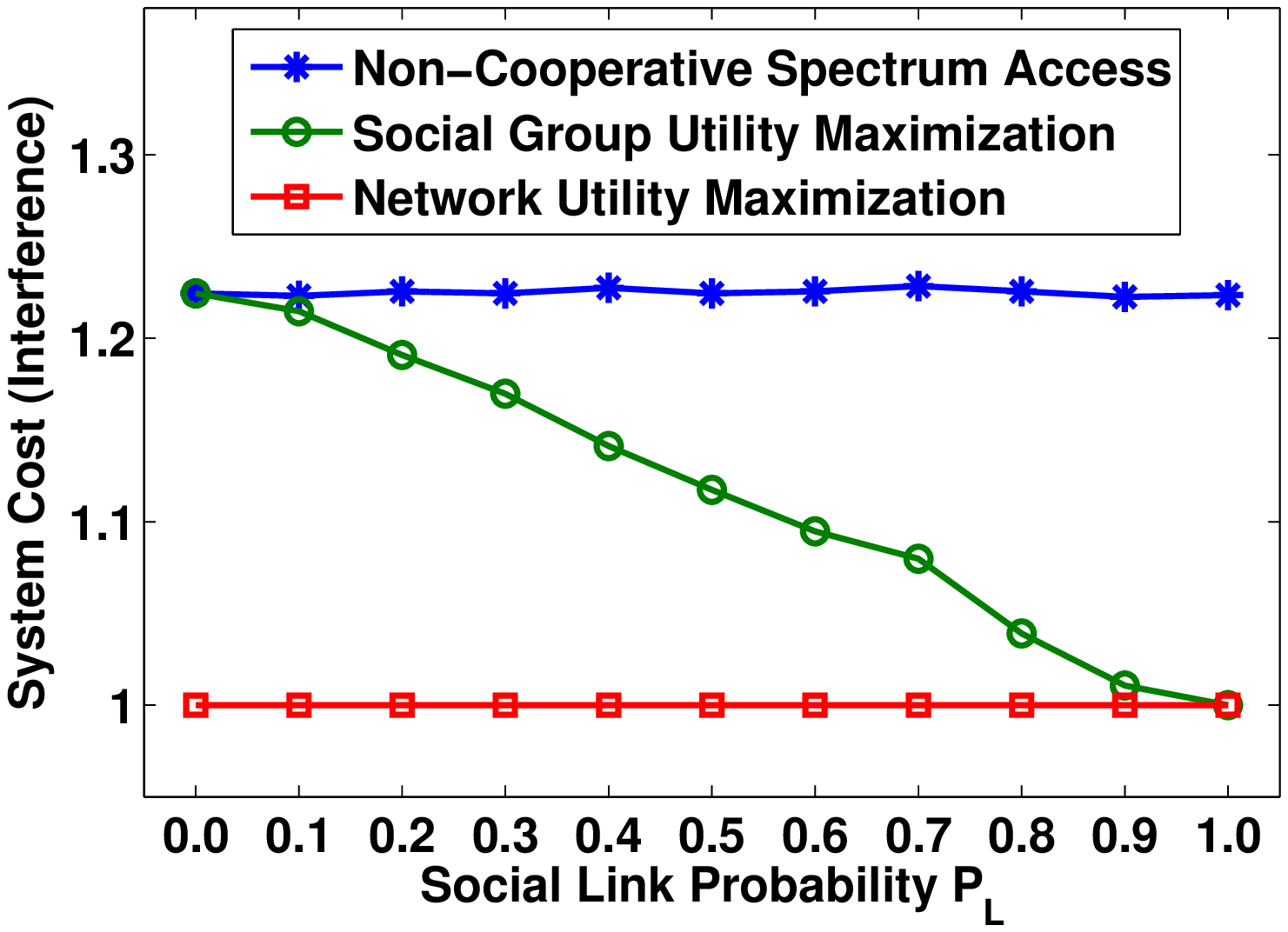}
\caption{\label{fig:System-wide-interference-with}Normalized system-wide interference
with different social network density}
\end{minipage}
\hfill
\begin{minipage}[t]{0.32\linewidth}
\centering
\includegraphics[scale=0.342]{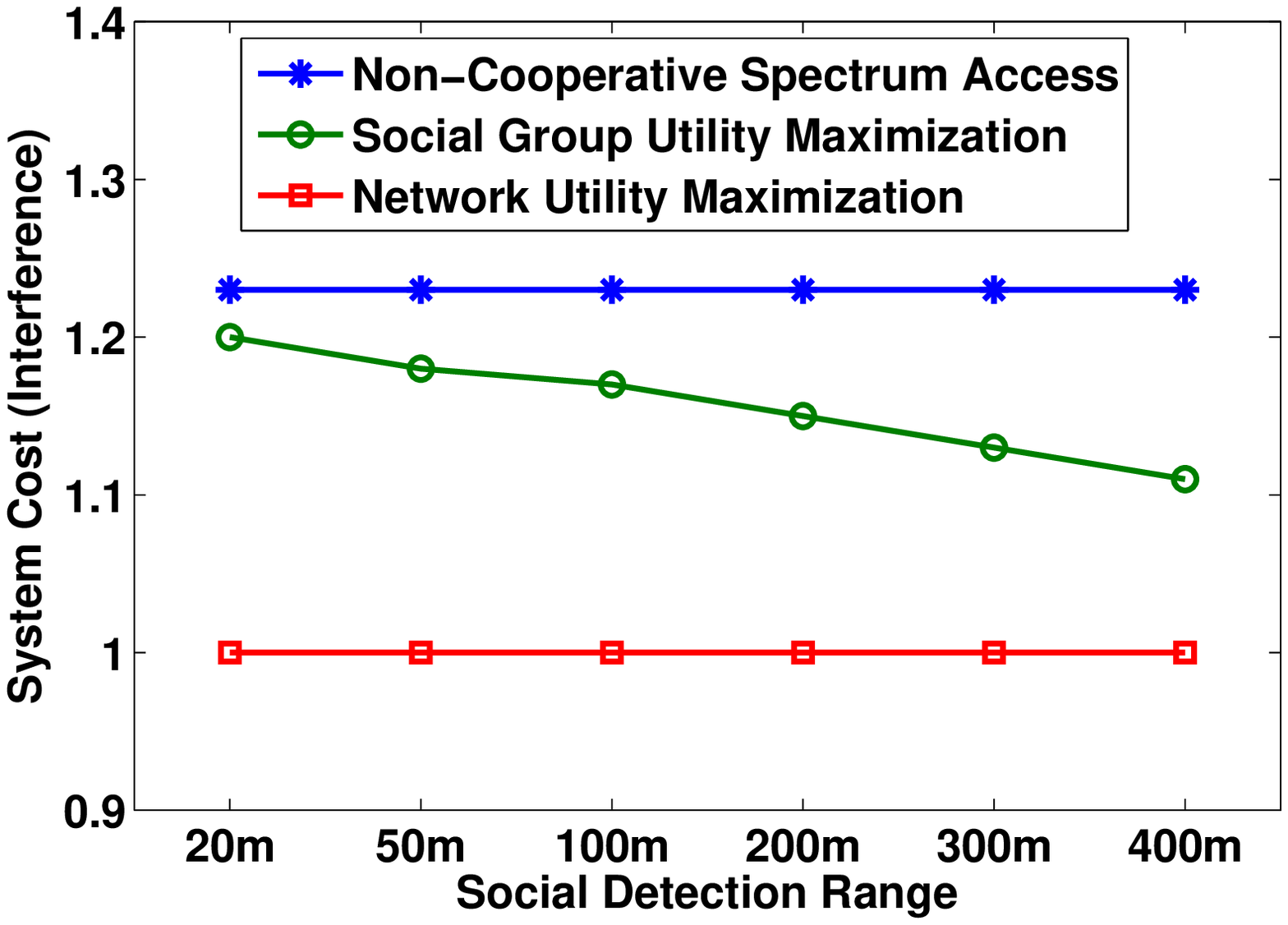}
\caption{\label{fig:range}Normalized system-wide interference
with different social detection ranges}
\end{minipage}
\hfill
\begin{minipage}[t]{0.32\linewidth}
\centering
\includegraphics[scale=0.37]{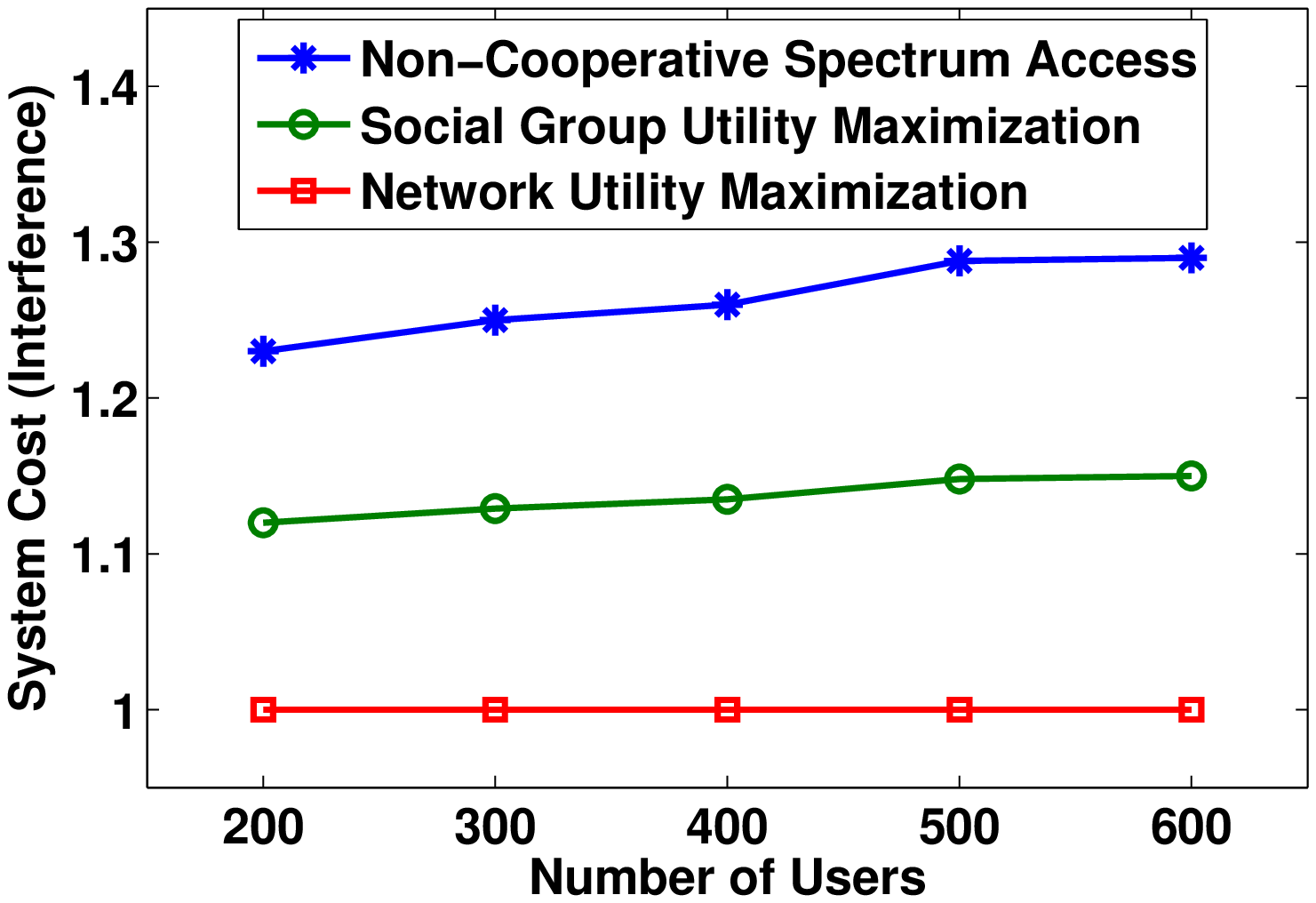}
\caption{\label{fig:System-wide-interference-with-1}Normalized system-wide interference
with different number of users}
\end{minipage}
\end{figure*}

%
%
%

We first consider a database assisted spectrum access network consisting
of $M=5$ channels and $N=8$ white-space users, which are scattered
across a square area of a length of $500$ m (see Figure \ref{fig:A-square-area}).
The transmission power of each user is $P_{n}=100$ mW \cite{FCC}, the path loss
factor $\alpha=4$, and the background interference power $\omega_{m}^{n}$
for each channel $m$ and user $n$ is randomly assigned in the interval
of $[-100,-90]$ dBm. Each user $n$ has a different set of
vacant channels by consulting the geo-location database. For example, the vacant channels for user $1$
are $\{2,3,4\}$. For the interference graph $\mathcal{G}^{p}$, we define that the user's transmission range $\delta=500$ m and two users can generate inference to each other if their
distance is not greater than $\delta$. The social
graph $\mathcal{G}^{s}$ is given in Figure \ref{fig:A-square-area} where two users have
social tie if there is an edge between them and the numerical value associated with each edge represents the strength of social tie.

We implement the proposed distributed spectrum access algorithm for
the SGUM game with different parameters $\theta$
in Figure \ref{fig:Potential-value-}. We see that the convergent
potential function value $\Phi$ of the SGUM game increases
as the parameter $\theta$ increases. When the parameter $\theta$
is large enough (e.g., $\theta\geq10^{6}$), the algorithm can approach
the maximum potential function value $\Phi^{*}=\max_{\boldsymbol{a}}\Phi(\boldsymbol{a})$.
To verify that the algorithm can approach the SNE
of the SGUM game, we show the dynamics of the potential
value $\Phi(\boldsymbol{a})$ in Figure \ref{fig:Dynamics-of-potential}.
We see that the distributed spectrum access algorithm can
drive the potential value $\Phi$ increasing and approach the maximum
potential value $\Phi^{*}$. According to the property of potential
game, the algorithm hence can approach the SNE of the SGUM game.

To verify the trade-off between the performance and the convergence time, we show in Figure \ref{fig:iteration}  the number iterations for convergence by the distributed spectrum access algorithm with different maximum allowable performance loss (by tuning the parameter $\theta$). We see that, compared with the case with the optimal performance, if we allow  at most $10\%$ and $20\%$ performance loss, then we can reduce the convergence time by up-to $69\%$ and $83\%$, respectively.

\subsubsection{Erdos-Renyi Social Graph}

We then consider $N=100$ users that randomly scattered across a square
area of a length of $2000$ m. We evaluate the SGUM game solution by the distributed spectrum access
algorithm with the social graph represented by the Erdos-Renyi
(ER) graph model \cite{newman2002random}, where a social link exists between any two
users with a probability of $P_{L}$. We set the strength
of social tie $w_{nm}=1$ for each social link. To evaluate the impact
of social link density of the social graph, we implement the simulations
with different social link probabilities $P_{L}=0,0.1,...,1.0$, respectively.
For each given $P_{L}$, we average over $100$ runs. To benchmark the SGUM solution, we also
implement the the following two solutions:

(1) Non-cooperative spectrum access: we implement the non-cooperative game based solution such that each user aims to maximize its individual utility, i.e., we set $S_{n}(\boldsymbol{a})=U_{n}(\boldsymbol{a})$ in the distributed spectrum access algorithm.

(2) Network utility maximization: we implement the social optimal solution such that the system-wide utility  is maximized, i.e., we set $S_{n}(\boldsymbol{a})=\sum_{n=1}^{N}U_n(\boldsymbol{a})$ in the distributed spectrum access algorithm.

Similar to the price of anarchy in non-cooperative game \cite{roughgarden2005selfish}, we normalize the system-wide interference of these solutions with respect to that of the social optimal solution (i.e., network utility maximum solution). The results are given in Figure \ref{fig:System-wide-interference-with}.
We see that the performance of the SGUM solution always
dominates that of the non-cooperative spectrum access. This is non-trivial since non-cooperative game promotes the competition among users to increase the system-wide utility and has been widely adopted to devise efficient distributed resource allocation mechanisms in wireless networks \cite{han2012game}. Moreover, we observe that
the performance gain of the SGUM solution increases
as the social link probability $P_{L}$ increases. When the social
link probability $P_{L}=1$, the SGUM solution achieves the same performance of the network utility maximization and can reduce $23\%$ system-wide interference over the non-cooperative spectrum access. This also demonstrates that the proposed SGUM framework spans the continuum between non-cooperative game and network utility maximization -- two extreme paradigms based on drastically different assumptions that users are selfish and altruistic, respectively.

We then investigate the scenario that a user may only take into nearby users with social ties. This can correspond to the case that a user can only detect the social friends within a fixed physical distance. In the simulation we set the social link probability $P_{L}=0.5$ and set the social detection range for identifying nearby social friends as $20m, 50m, ..., 400m$, respectively. We see from Figure \ref{fig:range} that the performance of SGUM solution improves as the social detection range increases. Intuitively, when the social detection range increases, the number of social links increases and hence the performance  of SGUM solution improves.

\subsubsection{Real Trace Based Social Graph}

We next evaluate the SGUM solution by
the distributed spectrum access algorithm based on the social
graph represented by the friendship network of the real data trace
Brightkite \cite{cho2011friendship}. Brightkite contains an explicit friendship network among the users. Different from the Erdos-Renyi (ER) social graph, the friendship network of Brightkite is scale-free such that the node degree distribution follows a power law \cite{KONECT2013}. We implement experiments with the number of users
$N=200,300,...,600,$ respectively. We randomly select $N$ nodes from Brightkite and construct the social graph based on the friendship relationship among these $N$ nodes in the friendship network of Brightkite. For each given $N$, we average over $1000$ runs.  As the benchmark, we also implement
the solutions of non-cooperative spectrum access and network utility maximization.

The results are shown in Figure \ref{fig:System-wide-interference-with-1}.
We see that the non-cooperative spectrum access solution will increase the system-wide interference up-to $29\%$ over the network utility maximization solution.  Upon comparison, the system-wide interference by the the SGUM solution will increase at most $15\%$, compared with the network utility maximization solution. This verifies the effectiveness of leveraging social tie to stimulate user cooperation for achieving efficient distributed spectrum access in practices.

\section{Social Group Utility Maximization Based Power Control}\label{sc:power}


In this section we study power control under the SGUM framework. Note that existing literature (e.g., \cite{Scutari06,Candogan10}) on power control game typically assume that users are selfish and do not take the social factors into account. In this section, we will investigate the impact of the social ties among the users on the system performance.

\subsection{SGUM Game Formulation}

We consider a set of users under the physical interference model, where user $i$ is a link consisting of a transmitter $T_i$ and a receiver $R_i$. The channel gain of communication link $i$ is $h_i$, and the channel gain of the interference link between transmitter $T_i$ and receiver $R_j$ is $g_{ij}$. The noise at receiver $R_i$ is $n_i$. Note that here we can view the physical graph $\mathcal{G}^{p}$ among users as a complete graph and the degree of physical coupling (i.e., interference) between any two users is captured by the channel gain.  Then the signal to interference and noise ratio (SINR) $\gamma_i$ of link $i$ is given by
\[\gamma_i(p_i,p_{-i}) = \frac{h_ip_i}{n_i + \sum^N_{j=1}g_{ji}p_j},\]
where $p_i$ denotes the transmit power of $T_i$. We consider that the individual utility $U_i$ of user $i$ is given by
\[U_i(p_i,p_{-i}) = \log(\gamma_j) - c_ip_i,\]
where $c_i$ denotes the cost of per unit power consumption. Similar to many studies \cite{Scutari06,Candogan10}, here we use the logarithmic function to model the utility of a user.

Then, given the social ties (i.e., $w_{ij}$) among the users on the social graph $\mathcal{G}^{s}$, we can define the social group utility function $S_i$  of user $i$ accordingly to (\ref{eq:SocialUtility0}). We then formulate the power control problem with social ties as a SGUM game $\Lambda \triangleq (\mathcal{N}, \{p_i\}, \{S_i\})$.

\subsection{SGUM Game Analysis}
For ease of exposition, we will first focus on the SGUM based power control game with two users, because the two-user case can shed light on the impact of social ties on users' strategies and social welfare.

\begin{thm}\label{thm:2user_NE}
For the two-user SGUM based power control game, there exists a unique SNE, where
\[p^{SNE}_1 = \sqrt{\alpha_1^2 + \beta_1} - \alpha_1, \ p^{SNE}_2 = \sqrt{\alpha_2^2 + \beta_2} - \alpha_2\]
with
\[\alpha_1 \triangleq \frac{w_{12}g_{12} + c_1n_2 - g_{12}}{2c_1g_{12}}, \ \beta_1 \triangleq \frac{n_2}{c_1g_{12}}\]
and
\[\alpha_2 \triangleq \frac{w_{21}g_{21} + c_2n_1 - g_{21}}{2c_2g_{21}}, \ \beta_2 \triangleq \frac{n_1}{c_2g_{21}}.\]
\end{thm}

The proof is given in the online appendix \cite{SGUM2016}. Using Theorem~\ref{thm:2user_NE}, we have the following result.

\begin{cor}\label{lm:2user_NE}
For the two-user SGUM based power control game, each user's SNE strategy decreases when its social tie with the other increases.
\end{cor}

\begin{figure}[t]
\centering
\includegraphics[width=0.35 \textwidth]{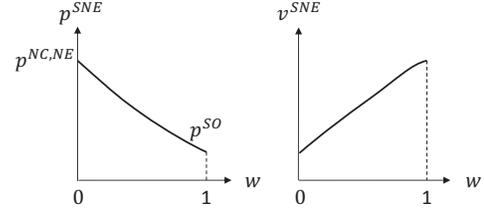}
\caption{For the two-user SGUM based power control game, as social tie $w\triangleq w_{12}=w_{21}$ increases from 0 to 1, each user's SNE strategy $p^{SNE}$ migrates from its NE strategy $p^{NC,NE}$ for a standard NCG to its social optimal strategy $p^{SO}$ for NUM. The social welfare of the SNE $v^{SNE}$ also migrates correspondingly.}
\label{fg:power_NE}
\end{figure}

Corollary~\ref{lm:2user_NE} shows that a users' SNE strategy is monotonically decreasing with the social tie. The following result shows that the social welfare of the SNE is also a monotonic function of social ties.

\begin{pp}\label{lm:2user_social}
For the two-user SGUM based power control game, the social welfare of the SNE increases when social ties increase.
\end{pp}

The proof is given in in the online appendix \cite{SGUM2016}. According to Proposition~\ref{lm:2user_social}, as illustrated in Figure~\ref{fg:power_NE}, when the social tie increases from $0$ to $1$, a user's SNE strategy migrates from its NE strategy for a standard NCG to its social optimal strategy for NUM.

We next study the existence of SNE for the SGUM based power control game with multiple users. Motivated by the observation that the two-user game is supermodular such that
\[\frac{\partial^2 S_1(p_1,p_{2})}{\partial p_1 \partial p_2} > 0 \mbox{ and } \frac{\partial^2 S_2(p_2,p_{1})}{\partial p_2 \partial p_1} > 0,\]
we have the following result.
\begin{thm}\label{thm:log_utility}
The SGUM based power control game with multiple users is a supermodular game, and thus there exists at least one SNE.
\end{thm}

The proof is given in the online appendix \cite{SGUM2016}. Since the SGUM game for power control is a supermodular game, similar to the scheme in~\cite{huang2006distributed},  users can start from the smallest transmission power (i.e., $p_i=0, \forall i\in\mathcal{N}$) and use asynchronous best response updates, so that their strategies will converge to a SNE. The game with more than two users does not yield closed-form SNE strategies, and hence is difficult to analyze the impact of social tie on system performance. In the following section we will use numerical results to show that the insight we draw from the analysis of
the two-user case also holds for the general multi-user case: the SNE for SGUM migrates monotonically from the NE for NCG to the social optimal solution for NUM.

\subsection{Numerical Results}

We consider $N$ users, each of which is a link consisting of a transmitter and a receiver. Each transmitter or receiver is randomly located in a square area with side length 500m. Under the physical interference model, we assume that the channel condition of a link (communication or interference link) only depends on the path loss effect with path loss factor $3$. We assume that the transmit power of each link is 1W and the noise power at each receiver is 0.1W. We simulate the
social graph based on the Erdos-Renyi (ER) graph, and the real data trace of
the friendship network from Brightkite.  We set the social tie of a social link as 1 if it exists.

\begin{figure}[tt]
\begin{minipage}[t]{0.48\linewidth}
\centering
\includegraphics[width=1.1\textwidth]{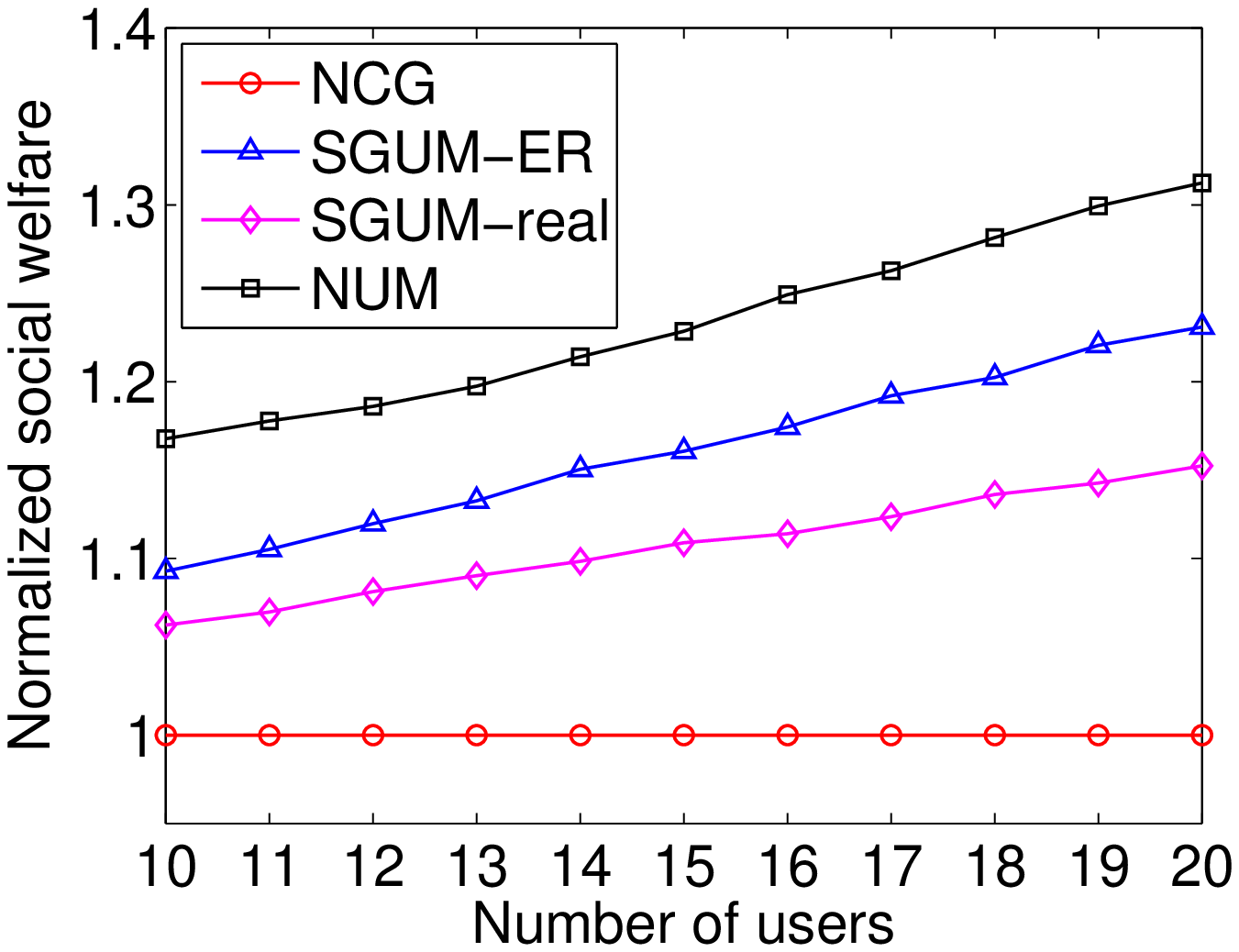}
\caption{Impact of $N$ for SGUM based power control game.}
\label{fg:power_N}
\end{minipage}
\hfill
\begin{minipage}[t]{0.48\linewidth}
\centering
\includegraphics[width=1.1\textwidth]{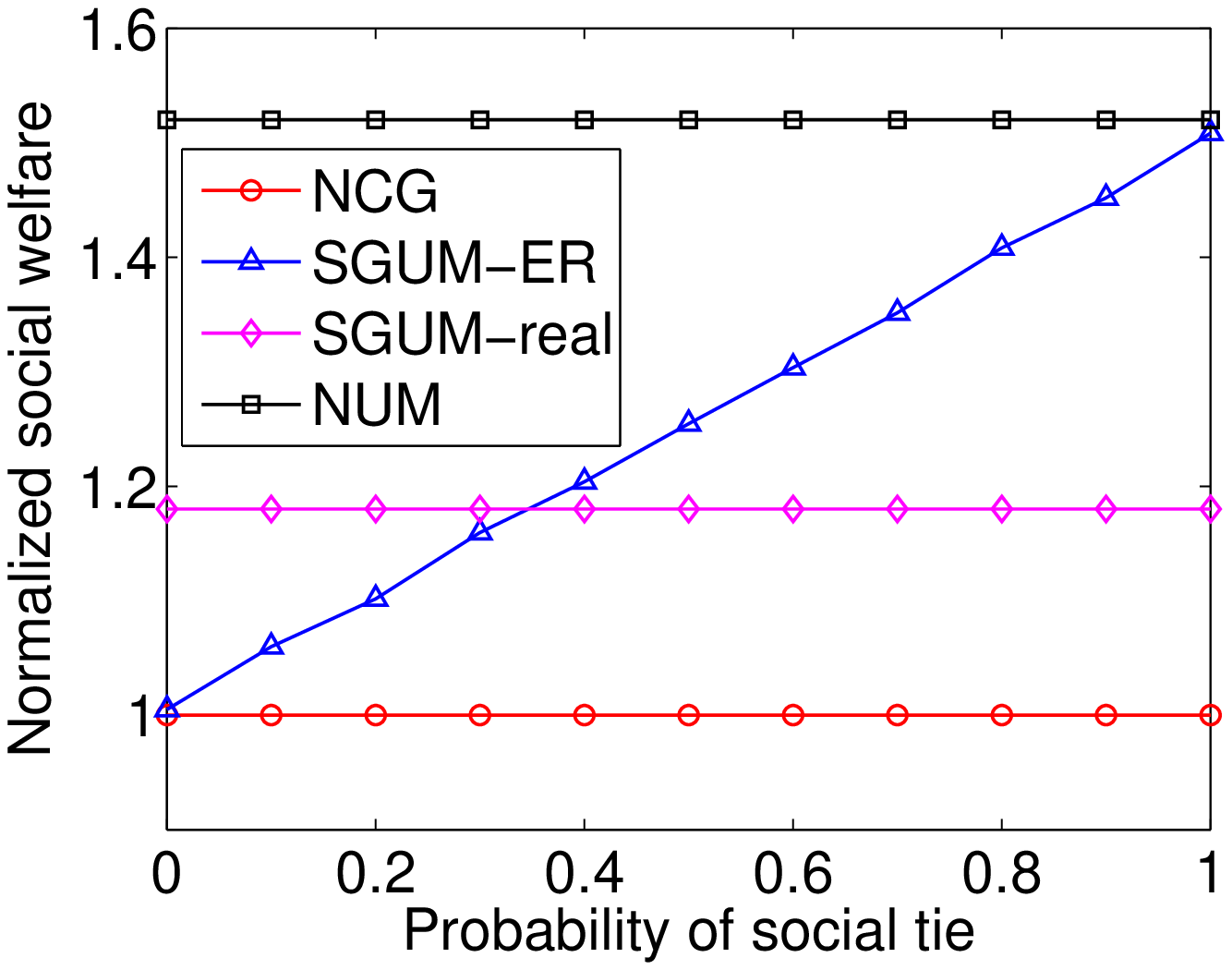}
\caption{Impact of $P_L$ for SGUM based power control game.}
\label{fg:power_Ps}
\end{minipage}
\end{figure}

Figure~\ref{fg:power_N} shows the normalized social welfare for a varying number of users $N$. For the Erdos-Renyi (ER) graph, the social link probability $P_L=0.5$. We can see that the SGUM solution for the ER model based social graph can achieve a performance gain up to 23\% over the NCG solution, and its performance loss from the NUM solution is at most 10\%. The SGUM solution for the real data based social graph can achieve a performance gain up to 15\%.

Figure~\ref{fg:power_Ps} shows the normalized social welfare as the social link density probability $P_L$  in the ER graph varies from 0 to 1. We set the number of users $N$ as 20. We observe that as $P_L$ increases, the social welfare of the SNE for the SGUM game increases monotonically from that of NE for NCG to that of the social optimal solution for NUM. This demonstrates that the SGUM spans the continuum between NCG and NUM.

%

\section{Social Group Utility Maximization Based Random Access Control}\label{sc:random}

In this section we study random access control under the SGUM framework. Note that most existing work (e.g., \cite{chen2010random,inaltekin2008analysis}) on random access control game assume that users are selfish and do not take the social factors into account. To best of our knowledge, \cite{kesidis2010stable} is the most related work, which considers a  two-user socially-aware random access control game with the assumption that the social tie among the two users are symmetric. While, we study the general case of multi-user socially-aware random access control game with asymmetric/sysmetric social ties among the users.

\subsection{SGUM Game Formulation}

We consider a set of users under the protocol interference model, where user $i$ is a link consisting of transmitter $T_i$ and receiver $R_i$. We can construct the the physical graph $\mathcal{G}^{p}$ among users for the random access control as follows. Let $\mathcal{I}^+_i$ denote the set of receivers that transmitter $T_i$ cause interference to, and $\mathcal{I}^-_i$ denote the set of transmitters that causes interference to receiver $R_i$. Each user $i$ contends for the opportunity of data transmission with probability $q_i\in[0,1]$ in a time slot. If multiple interfering links contend in the same time slot, a collision occurs and no link can grab the transmission opportunity. Then the probability $b_i$ that user $i$ can grab the transmission opportunity is given by
\begin{align*}
b_i(q_i, q_{-i}) = q_i\prod_{j\in\mathcal{I}^-_i}(1 - q_j).
\end{align*}
We assume that the individual utility of user $i$ is given by
\begin{align*}
U_i(q_i,q_{-i}) = \log(z_ib_i(q_i, q_{-i})) - c_iq_i,
\end{align*}
where $z_i>0$ represents user $i$'s efficiency of utilizing the transmission opportunity (e.g., transmission rate), and $c_i>0$ represents user $i$'s cost of contention (e.g., energy consumption).

Then, given the social ties (i.e., $w_{ij}$) among the users on the social graph $\mathcal{G}^{s}$, we can define the social group utility function $S_i$ of user $i$ accordingly to (\ref{eq:SocialUtility0}). We then formulate the random access control problem with social ties as a SGUM game $\Xi \triangleq (\mathcal{N}, \{q_i\}, \{S_i\})$.


\subsection{SGUM Game Analysis}

For the SGUM based random access control game, we have the following result.

\begin{thm}\label{thm:random_NE}
For the SGUM based random access control game, there exists a unique SNE, where for any user $i\in \mathcal{N}$, the access probability  $q_i^{SNE}$ is
\begin{align}\label{SNE_random}
 \frac{\sum_{j\in\mathcal{I}^+_i} w_{ij}+1+c_i-\sqrt{(\sum_{j\in\mathcal{I}^+_i} w_{ij}+1+c_i)^2-4c_i}}{2c_i}.
\end{align}
\end{thm}


The proof is given in the online appendix \cite{SGUM2016}. Using Theorem~\ref{thm:random_NE}, by showing the first-order derivative of the access probability $q_i^{SNE}$ in terms of $w_{ij}$ is negative, we have the following result.

\begin{cor}\label{lm:random_NE}
For the SGUM based random access control game, each user's SNE strategy decreases when its social ties with others increase.
\end{cor}

\begin{figure}[t]
\centering
\includegraphics[width=0.35 \textwidth]{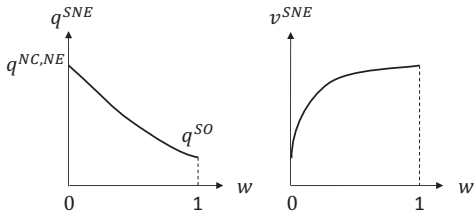}
\caption{For a two-user SGUM based random access control game, as social tie $w\triangleq w_{12}=w_{21}$ increases from 0 to 1, each user's SNE strategy $q^{SNE}$ migrates from its NE strategy $q^{NC,NE}$ for a standard NCG to its socially optimal strategy $q^{SO}$ for NUM. The social welfare of the SNE $v^{SNE}$ also migrates correspondingly.}
\label{fg:random_NE}
\end{figure}

Corollary~\ref{lm:random_NE} shows that a users' SNE strategy is monotonically decreasing with the social tie. Intuitively, when a user's access probability increases, it increases the collision probability of the users within its interference range, and hence reduces their individual utilities. Therefore, a user would reduce its access probability when its social ties with those within its interference range increase (as illustrated in Figure~\ref{fg:random_NE}). Furthermore, Theorem~\ref{thm:random_NE} shows that each user's SNE strategy is independent of other users' strategies, but depends on the user's social ties with others. This can facilitate the implementation of the SGUM game solution for random access control in a distributed manner, such that each user $i$ locally sets the access probability accordingly to the social ties of its neighbors.

The following result shows that the social welfare of the SNE is also a monotonic function of social ties.

\begin{pp}\label{lm:random_social}
For the SGUM based random access control game, the social welfare of the SNE increases when social ties increase.
\end{pp}

The proof is given in the online appendix \cite{SGUM2016}. Intuitively, since users' individual utilities are summed up with equal weights in the social welfare, a user's SNE strategy becomes closer to its social optimal strategy when other users weigh more (i.e., the social ties increase) in that user's social group utility, and the social welfare also increases. As illustrated in Figure~\ref{fg:random_NE}, when social ties increase, a user's SNE strategy migrates from its NE strategy for a standard NCG to its socially optimal strategy for NUM. This demonstrates that the SGUM game framework spans the continuum between these traditionally disjoint paradigms.

\subsection{Numerical Results}
We consider $N$ users, each of which is a link consisting of a transmitter and a receiver. Each transmitter or receiver is randomly located in a square area with side length 500m. Under the
protocol interference model, we assume that a link causes interference to another link if the former link's transmitter is within 100m of the latter link's receiver. We simulate the
social graph based on the Erdos-Renyi (ER) graph with the social link probability $P_L=0.5$, and the real data trace of
the friendship network from Brightkite.  We set the social tie of a social link as 1 if it exists.

\begin{figure}[t]
\centering
\includegraphics[width=0.25\textwidth]{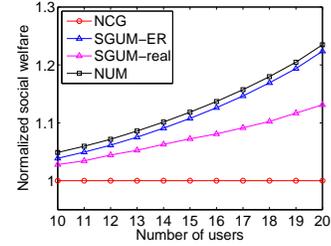}
\caption{Impact of $N$ for SGUM based random access control game.}
\label{fg:random_N}
\end{figure}

To illustrate the system efficiency of the SGUM solution, we compare it with the NCG solution where each user aims to maximize its individual utility, and the NUM solution where the total
individual utility of all users is maximized. Figure~\ref{fg:random_N} depicts the social welfare of the SNE for SGUM and the social optimal solution for NUM normalized with respect to the NE
for NCG, as the number of users increases. We can see that the SGUM solution for the ER model based social graph always dominates that of the NCG, with a substantial performance gain up to
22\%. On the other hand, it performs almost as well as the NUM solution. This demonstrates that system efficiency can be significantly improved by exploiting social ties. We observe that the
SGUM solution for the real data based social graph is worse than that for the ER model based social graph due to that social ties are weaker in the real data than in the ER graph with link
probability $0.5$. However, it still can achieve a performance gain up to 13\% over that of the NCG solution.

\section{Further Discussion}\label{extension}

In the sections above, we study the social group utility maximization (SGUM) framework with the ``positive" social tie between two users (e.g., family members or friends). In this case, one user cares about the welfare of the other. To generalize  the SGUM framework, we can further consider the case that the social tie between two users is ``negative" (e.g., due to malicious behavior) such that one user may intend to damage the other's welfare. Hence, the generalized SGUM framework with both positive and negative social ties can be very useful for modeling the network security issues.

%
%

\begin{figure}[t]
\centering
\includegraphics[width=0.35\textwidth]{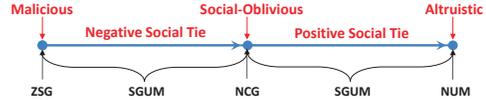}
\caption{The social group utility maximization (SGUM) game framework spans the continuum from zero-sum game (ZSG) to general-sum non-cooperative game (NCG) to network utility maximization (NUM).}
\label{fg:bridge_gap}
\end{figure}

Formally, we can define that $w_{ij}\in(-\infty,1]$: When $w_{ij}\in(0,1]$, it represents the extent to which user $i$ cares about user $j$'s utility, and it reaches the highest when $w_{ij}=1$ (i.e., user $i$ cares about user $j$'s utility as much as its own utility); when $w_{ij}\in(-\infty,0)$, it pinpoints to how much user $i$ intends to damage user $j$'s utility, and reaches the extreme as $w_{ij}$ goes to $-\infty$ (i.e., user $i$ would sacrifice all of its own utility to damage user $j$'s utility).

Zero-sum game (i.e., all players' payoffs are summed up to $0$) is widely adopted for modelling the interactions between attackers and defenders in network security problems \cite{ibidunmoye2013modeling}. Interestingly, the generalized SGUM framework also encompasses the zero-sum game as a special case. Specifically, when the accumulated negative social ties of each user satisfies that $\sum_{j:i\in\mathcal{N}^{s}_{j}}w_{ji}=-1$, for any $i\in\mathcal{N}$, we have that
\begin{align*}
\sum_{i=1}^{N}S_{i}(\boldsymbol{x})  =   \sum_{i=1}^{N}U_{i}(\boldsymbol{x})+\sum_{i=1}^{N}U_{i}(\boldsymbol{x})\sum_{j:i\in\mathcal{N}_{j}^{s}}w_{ji}=0.
\end{align*}
In this case, the SGUM game degenerates to a zero-sum game, where each user views the total gain of other users as its loss (see Figure \ref{fg:bridge_gap} for an illustration). For example, a SGUM game of two users with $S_1=U_1-U_2$ and $S_2=U_2-U_1$, or $S_1=U_1$ and $S_2=-U_1$, is a zero-sum game.

Building on that the generalized SGUM framework with both positive and negative social ties,  we can devise more effective and efficient defense mechanisms against the attacks by the malicious users (i.e., users with negative social ties), by leveraging trustworthy helps and collaborations from the social friends (i.e., users with positive social ties). Due to space limit, we will pursuit a thorough understanding of the generalized SGUM framework in a future work.

\section{\label{conclusion}Conclusion}
In this paper, we have developed a general SGUM framework that highlights the interplay between the physical coupling and the social coupling among users. We showed that the SGUM framework can provide rich modeling flexibility and span the continuum between non-cooperative game and network utility maximization. In particular, we have studied the application in database assisted spectrum access, power control, and random access control under this framework. For the case of database assisted spectrum access, we show that the SGUM game is a potential game and always admits a socially-aware Nash equilibrium (SNE). We also design a distributed spectrum access algorithm that can achieve the SNE of the game and quantify its performance gap. We further show that the upper-bound of the performance gap decreases as the strength of social ties among users increases.  For the cases of power control and random access control, we showed that there exists a unique SNE. Furthermore, as the strength of social ties increases from the minimum to the maximum, a player's SNE strategy (i.e.,  transmit power or access probability) migrates from the Nash equilibrium strategy in a standard non-cooperative game to the socially-optimal strategy in network utility maximization.

We are currently studying the SGUM framework with both positive and negative social ties to build a thorough understanding of the impact of altruistic and malicious behaviors on network performance and security. Beyond the supermodular/potential game approaches in this paper, one can resort to other powerful tools such as \emph{monotone game} to characterize SNE and design distributed mechanisms accordingly.  We believe that this framework will open a new door for future networking system design by exploiting social interactions.

\bibliographystyle{ieeetran}
\bibliography{SocialSpectrum}

\end{document}